\pdfoutput=1
\documentclass[12pt]{cernprep}
\usepackage{epsfig,color,rotating}
\usepackage{lscape}
\begin{document}
\newcommand{\dedx}{\mbox{${\rm d}E/{\rm d}x$}}
\newcommand{\EcB}{$E \! \times \! B$}
\newcommand{\omt}{$\omega \tau$}
\newcommand{\omtsq}{$(\omega \tau )^2$}
\newcommand{\rphi}{\mbox{$r \! \cdot \! \phi$}}
\newcommand{\srphi}{\mbox{$\sigma_{r \! \cdot \! \phi}$}}
\newcommand{\dg}{\mbox{`durchgriff'}}
\newcommand{\mg}{\mbox{`margaritka'}}
\newcommand{\pT}{\mbox{$p_{\rm T}$}}
\newcommand{\GeVc}{\mbox{GeV/{\it c}}}
\newcommand{\MeVc}{\mbox{MeV/{\it c}}}
\def\kr{$^{83{\rm m}}$Kr\ }
\begin{titlepage}
\date{14 March 2008} 
%
%
\vspace{1cm}
\title{COMPARISON OF GEANT4 HADRON GENERATION WITH DATA \\
FROM THE INTERACTIONS WITH BERYLLIUM NUCLEI OF 
\mbox{\boldmath $+8.9$}~GeV/\mbox{\boldmath $c$} 
PROTONS AND PIONS,
AND OF \mbox{\boldmath $-8.0$}~GeV/\mbox{\boldmath $c$} 
PIONS}

\begin{abstract}
Hadron generation in the Geant4 simulation tool kit 
is compared with inclusive spectra of secondary protons 
and pions from the interactions with beryllium nuclei of $+8.9$~GeV/{\it c} 
protons and pions, and of $-8.0$~GeV/{\it c} pions. The data were taken in 2002 
at the CERN Proton Synchrotron with the HARP spectrometer.
We report on significant disagreements between data and 
simulated data especially in the polar-angle distributions of secondary protons and pions.
\end{abstract}

\vfill  \normalsize
\begin{center}
The HARP--CDP group  \\  

\vfill

\begin{Authlist}

A.~Bolshakova, I.~Boyko, G.~Chelkov,    
D.~Dedovitch, A.~Elagin$^1$, M.~Gostkin, A.~Guskov, 
Z.~Kroumchtein, Yu.~Nefedov, 
K.~Nikolaev, A.~Zhemchugov
\Instfoot{a1}{{\bf Joint Institute for Nuclear Research, Dubna, Russia}}
F.~Dydak, J.~Wotschack$^*$
\Instfoot{a2}{{\bf CERN, Geneva, Switzerland}}
A.~De~Min$^2$
\Instfoot{a3}{{\bf Politecnico di Milano and INFN, Sezione di
Milano-Bicocca, Italy}}
V.~Ammosov, V.~Gapienko, V.~Koreshev, A.~Semak, Yu.~Sviridov, E.~Usenko$^3$, V.~Zaets
\Instfoot{a4}{{\bf Institute for High Energy Physics, Protvino, Russia}}
\end{Authlist}

\vspace*{5mm}

\submitted{(To be submitted to Eur. Phys. J. C)}
\end{center}

\vspace*{5mm}
\rule{0.9\textwidth}{0.2mm}

\begin{footnotesize}
$^1$~Now at Texas A\&M University, College Station, USA 

$^2$~On leave of absence at 
Ecole Polytechnique F\'{e}d\'{e}rale, Lausanne, Switzerland 

$^3$~Now at Institute for Nuclear Research RAS, Moscow, Russia.

$^*$~Corresponding author; e-mail: joerg.wotschack@cern.ch
\end{footnotesize}

\end{titlepage}

\newpage

\vspace{0.8cm}

\newpage 

\section{Introduction and motivation}

The HARP experiment arose from the realization that the 
differential cross-sections of hadron production 
in the collisions of few GeV/{\it c} protons with nuclei were 
known only within a factor of two to three. Consequently, 
the HARP spectrometer was designed to carry out a programme 
of systematic and precise measurements of hadron production 
by protons and pions with momenta from 3 to 15~GeV/{\it c}. 
The experiment was in operation at the CERN Proton Synchrotron 
in 2001 and 2002, with a set of stationary targets ranging 
from hydrogen to lead, including beryllium.

The data from the HARP spectrometer 
can be used, amongst other purposes, for the physics validation of 
hadron generators that are used in 
simulation tool kits such as 
Geant4~\cite{Geant4}. This is of interest for the
correct interpretation of data that will
be forthcoming, e.g., from experiments at the 
LHC~\cite{DeRoeck}. 

In this paper, data are used from the HARP large-angle spectrometer 
that comprised a cylindrical Time Projection 
Chamber (TPC) and an array of Resistive Plate Chambers (RPCs) 
around the TPC. The purpose of the TPC was the measurement
of the transverse momentum $p_{\rm T}$ and of the polar angle
$\theta$ of tracks, and particle identification by d$E$/d$x$. The 
purpose of the RPCs was a complementary particle identification
by time of flight.

The data analysis that underlies the spectra shown 
in this paper rests on the calibrations of the 
TPC and the RPCs that our group published in 
Refs.~\cite{TPCpub} and \cite{RPCpub}. For a 
more detailed account of our calibration work we refer to our 
collection of memos and analysis notes~\cite{HARPCDPbibliography}.  
We recall that we disagree with the calibrations 
and physics results reported by the `HARP Collaboration', 
as discussed in Refs.~\cite{JINSTpub} and \cite{EPJCpub}. 

With a view to correcting for losses of secondary particles
from acceptance cuts, and for migration due to finite detector 
resolution, the measurement of tracks in the detector must
be simulated with a Monte Carlo program.
We use the Geant4 tool kit for this purpose. 
It was at this point that we noticed peculiar structures
in the polar-angle spectra of secondary particles 
generated by Geant4's LHEP `physics list' 
that prevented the weighting of generated tracks by
smooth functions. Further investigations showed that
this is a rather common phenomenon across  
Geant4's hadronic physics lists. 

\begin{figure}[htp]
\begin{center}
\includegraphics[width=9cm]{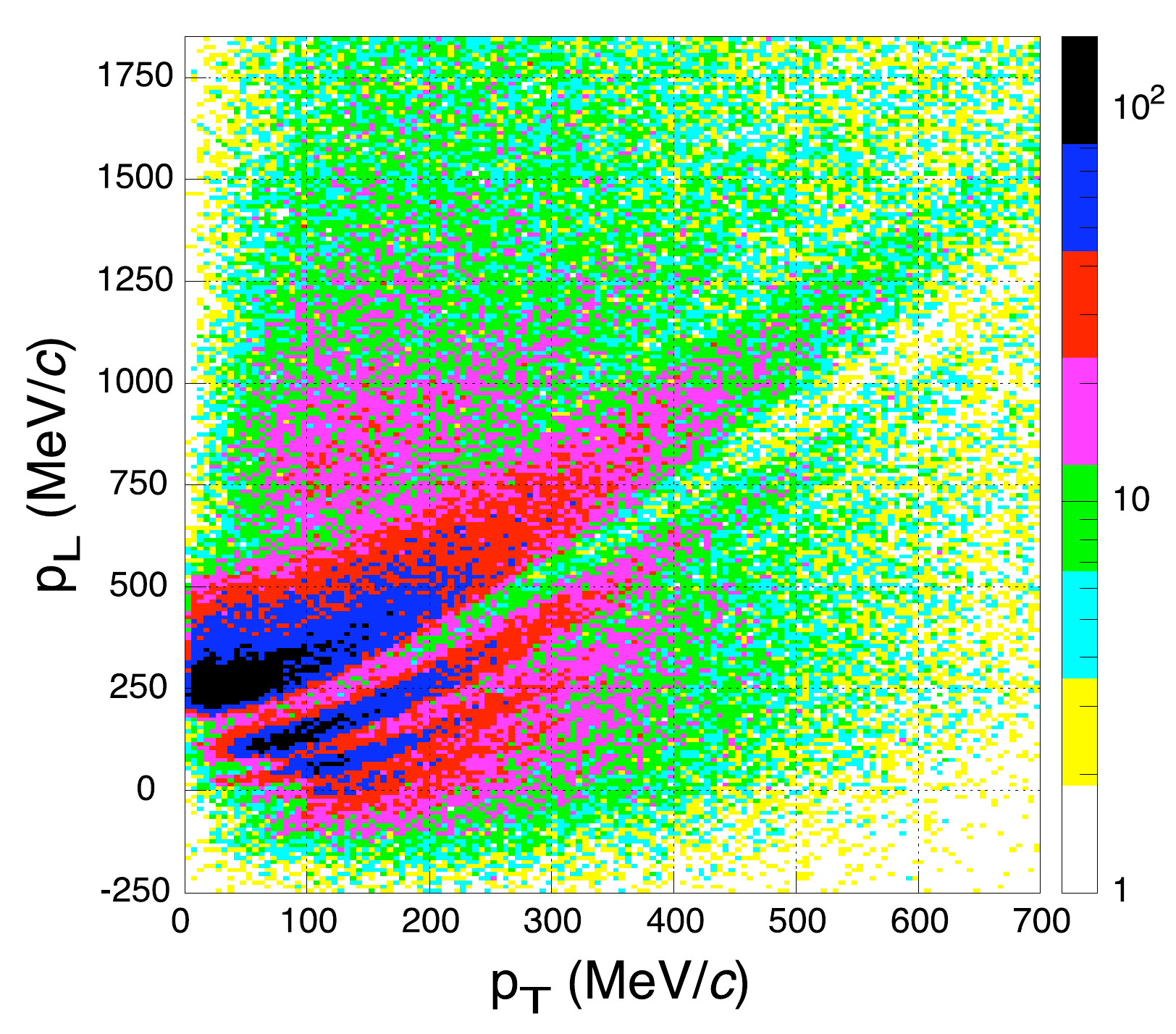} 
\caption{Longitudinal momentum $p_{\rm L}$ versus
transverse momentum $p_{\rm T}$, as generated by Geant4's  
LHEP physics list for secondary $\pi^+$ from the interactions
of $+8.9$~GeV/{\it c} beam $\pi^+$ with beryllium nuclei
at rest.}
\label{zebraplot}
\end{center}
\end{figure}

Figure~\ref{zebraplot} shows a
typical example of an unphysical structure in 
generated longitudinal momentum $p_{\rm L}$ versus transverse
momentum $p_{\rm T}$ of secondaries. 
Since the structure
is genuinely connected with the polar angle $\theta$, 
it tends to be washed out when
integrating over either 
$p_{\rm L}$ or $p_{\rm T}$\footnote{All physical quantities 
in this paper refer to the laboratory system.}.
That may explain why---as nearly as we can tell---these
structures were not noticed before.

\section{Hadron generators in the Geant4 simulation tool kit}
\label{generators}

The Geant4 simulation tool kit provides several 
physics models of hadronic interactions of hadrons with nuclei, 
and collections of such models, termed physics lists. 
The latter are tailored with a view to optimizing 
performance for specific applications.  

Table~\ref{physicslistsoverview} lists and characterizes 
a representative selection of physics lists 
of hadronic interactions 
in Geant4\footnote{Version 9.1 dated 14 December, 2007.}, 
together with the used physics models and the energy ranges 
where the latter are considered to be reliable~\cite{Geant4RefMan}. 

In the so-called `low-energy' domain
(defined as kinetic energy 
$E$ of the incoming hadron below some $25$~GeV), a modified version
of the GHEISHA package of Geant3 is used in many physics lists: 
the Parametrized Low-Energy Model (`LE\_GHEISHA'). Optionally, 
for $E$ below a few GeV, the 
Bertini Cascade~\cite{bertini} (`BERT') or the 
Binary Cascade~\cite{binary} (`BIC') 
models can be enabled, with a view to simulating the cascading
of final-state hadrons when they move through nuclear matter.  
As an alternative to LE\_GHEISHA, a modified 
version of the FRITIOF string fragmentation 
model~\cite{fritiof} (`FTF') is available. 

In the so-called `high-energy' domain, mostly the Quark--Gluon String Model
(`QGSM') is used, with FTF and the Parametrized High-Energy Model
(`HE\_GHEISHA') as alternatives. Further terms that appear in
Table~\ref{physicslistsoverview} and are explained in 
Ref.~\cite{Geant4RefMan}, are `PRECO' for the 
Pre-compound model, `QEL' for the Quasi-elastic scattering model,
and `CHIPS' for the Chiral Invariant Phase Space model.  

The energy ranges of models tend to overlap. In the overlap region,
the model is chosen randomly but the choice is biased by the
difference between the kinetic energy of the beam particle and
the kinetic energy limits of the models.

\begin{table}[htp]
\caption{Overview of selected physics lists 
of hadronic interactions in Geant4.}
\label{physicslistsoverview}
\begin{center}
\small
\begin{tabular}{|c|c|c|c|c|} \hline
Physics list & \multicolumn{2}{c|}{Proton beam} 
             & \multicolumn{2}{c|}{$\pi^\pm$ beam } \\ 
\hline
\hline 
LHEP & HE\_GHEISHA   & 25 GeV--100 TeV  & HE\_GHEISHA & 25 GeV--100 TeV \\
     & LE\_GHEISHA & 0--55 GeV & LE\_GHEISHA  &  0--55 GeV \\ 
\hline
LHEP\_PRECO\_HP & HE\_GHEISHA & 25 GeV--100 TeV & HE\_GHEISHA 
     & 25 GeV--100 TeV \\
     & LE\_GHEISHA & 0.15--55 GeV & LE\_GHEISHA & 0--55 GeV \\
     & PRECO & 0--0.17 GeV &            &        \\
\hline
QGSC & QGSM+QEL+CHIPS & 8 GeV--100 TeV & QGSM+QEL+CHIPS & 8 GeV--100 TeV \\
     & LE\_GHEISHA & 0--25 GeV & LE\_GHEISHA & 0--25 GeV \\ 
\hline
QGS\_BIC & QGSM+BIC & 12 GeV--100 TeV & QGSM+QEL & 12 GeV--100 TeV \\
     & LE\_GHEISHA & 9.5--25 GeV & LE\_GHEISHA & 1.2--25 GeV \\
     & BIC & 0--9.9 GeV & BIC & 0--1.3 GeV \\
\hline 
QGSP & QGSM+QEL+PRECO & 8 GeV--100 TeV 
     & QGSM+QEL+PRECO & 8 GeV--100 TeV \\
     & LE\_GHEISHA & 0--25 GeV & LE\_GHEISHA & 0--25 GeV \\
\hline
QGSP\_BERT & QGSM+QEL+PRECO & 8 GeV--100 TeV 
     & QGSM+QEL+PRECO & 8 GeV--100 TeV \\
     & LE\_GHEISHA & 9.5--25 GeV & LE\_GHEISHA & 9.5--25 GeV \\ 
     & BERT    & 0--9.9 GeV & BERT & 0--9.9 GeV \\      
\hline
QGSP\_BIC & QGSM+QEL+PRECO & 8 GeV--100 TeV 
     & QGSM+QEL+PRECO & 8 GeV--100 TeV \\
     & LE\_GHEISHA & 9.5--25 GeV & LE\_GHEISHA & 0--25 GeV \\ 
     & BIC & 0--9.9 GeV  &  &  \\ 
\hline 
QBBC & QGSM+QEL+CHIPS & 6 GeV--100 TeV & QGSM+QEL+CHIPS & 6 GeV--100 TeV \\
     & BIC  & 0--9 GeV & BERT        & 0--9 GeV \\ 
\hline
FTFC & FTF+QEL+CHIPS  & 4 GeV--100 TeV &  FTF+QEL+CHIPS & 4 GeV--100 TeV \\
     & LE\_GHEISHA & 0--5 GeV & LE\_GHEISHA & 0--5 GeV \\ 
\hline
FTFP & FTF  & 4 GeV--100 TeV & FTF+QEL+PRECO  &  4 GeV--100 TeV \\
     & LE\_GHEISHA & 0--5 GeV & LE\_GHEISHA & 0--5 GeV \\
 \hline
FTFP\_BERT & FTF  & 4 GeV--100 TeV & FTF+QEL+PRECO & 4 GeV--100 TeV \\
     & BERT & 0-5 GeV & BERT & 0-5 GeV \\ 
\hline
\end{tabular}
\end{center}
\normalsize
\end{table}

Below, we compare the predictions of Geant4 hadronic physics lists with our data: the inclusive 
proton, $\pi^+$ and $\pi^-$ spectra that are generated by the interactions with beryllium nuclei of $+8.9$~GeV/{\it c} protons 
and $\pi^+$, and of $-8.0$~GeV/{\it c} $\pi^-$.

\section{The HARP large-angle spectrometer}

\subsection{Physics performance}

For the purpose of this paper, the essential physics performance parameters are the resolution and the scale of the transverse momentum $p_{\rm T}$ of final-state particles, the resolution and the scale of the polar angle $\theta$, and the separation of pions from protons. We briefly give evidence of the salient features, and refer the reader to our respective technical publications~\cite{TPCpub,RPCpub} for details.

\begin{figure}[htp]
\begin{center}
\includegraphics[width=9cm]{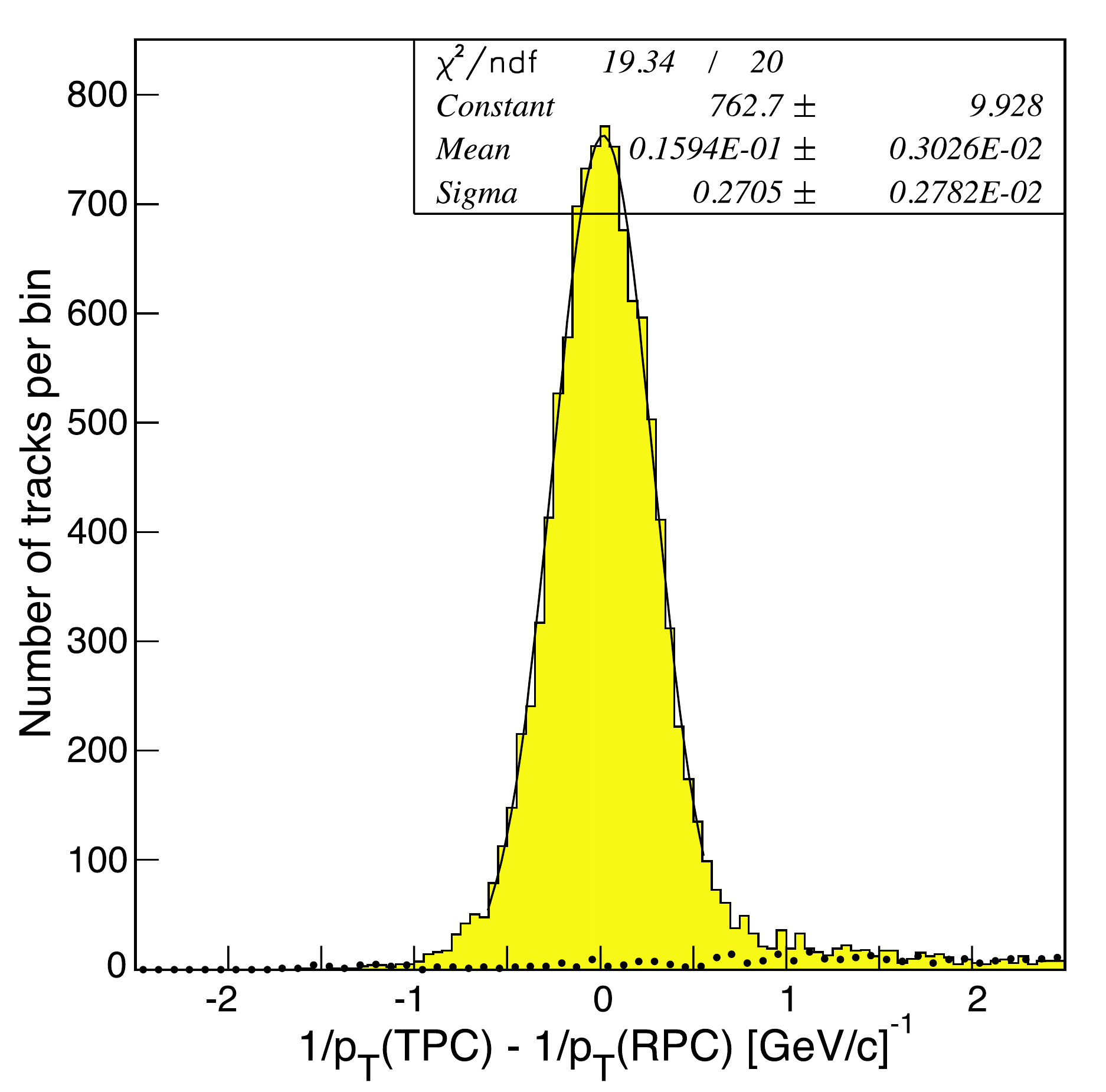} 
\caption{Difference of the inverse transverse momenta of positive
(shaded histogram) and negative (black points) particles 
from the measurement in the TPC and from the determination from 
RPC time of flight, for $0.6 < \beta < 0.75$ and 
for $45^\circ < \theta < 65^\circ$; the positive
particles are protons, with a very small background from pions
and kaons.}
\label{pTresolution}
\end{center}
\end{figure}

The resolution of the inverse transverse momentum measured by
the TPC depends slightly on the 
relative velocity $\beta$ and on $\theta$ of the 
particles. It is in the range
$0.20 < \sigma (1/p_{\rm T}) < 0.25$~(GeV/{\it c})$^{-1}$.
Figure~\ref{pTresolution} shows the difference of 
the inverse transverse momentum of positive particles with
$0.6 < \beta < 0.75$  
and $45^\circ < \theta < 65^\circ$ 
from the measurement in the TPC and from the
determination from RPC time of flight with the proton-mass
hypothesis. The positive particles are protons, the 
background from pions and kaons is very
small. Subtracting quadratically from the convoluted 
resolution of 0.27~(GeV/{\it c})$^{-1}$ the  
contribution from the 
time-of-flight resolution of the RPC,  
gives a net TPC resolution of
$\sigma (1/p_{\rm T}) = 0.20$~(GeV/{\it c})$^{-1}$.

From the requirement that $\pi^+$ and $\pi^-$ with the same RPC time of flight have the same momentum, the momentum scale is
determined to be correct to better than 2\%, for both positively and negatively charged particles.

The polar angle $\theta$ is measured in the TPC with a 
resolution of $\sim$9~mrad, for a representative 
angle of $\theta = 60^\circ$. To this a multiple scattering
error has to be added which is $\sim$7~mrad for a proton with 
$p_{\rm T} = 500$~MeV/{\it c} and $\theta = 60^\circ$, and 
$\sim$4~mrad for a pion with the same characteristics.
The polar-angle scale is correct to better than 2~mrad.     

\begin{figure}[h]
\begin{center}
\begin{tabular}{cc}
\includegraphics[width=7.6cm]{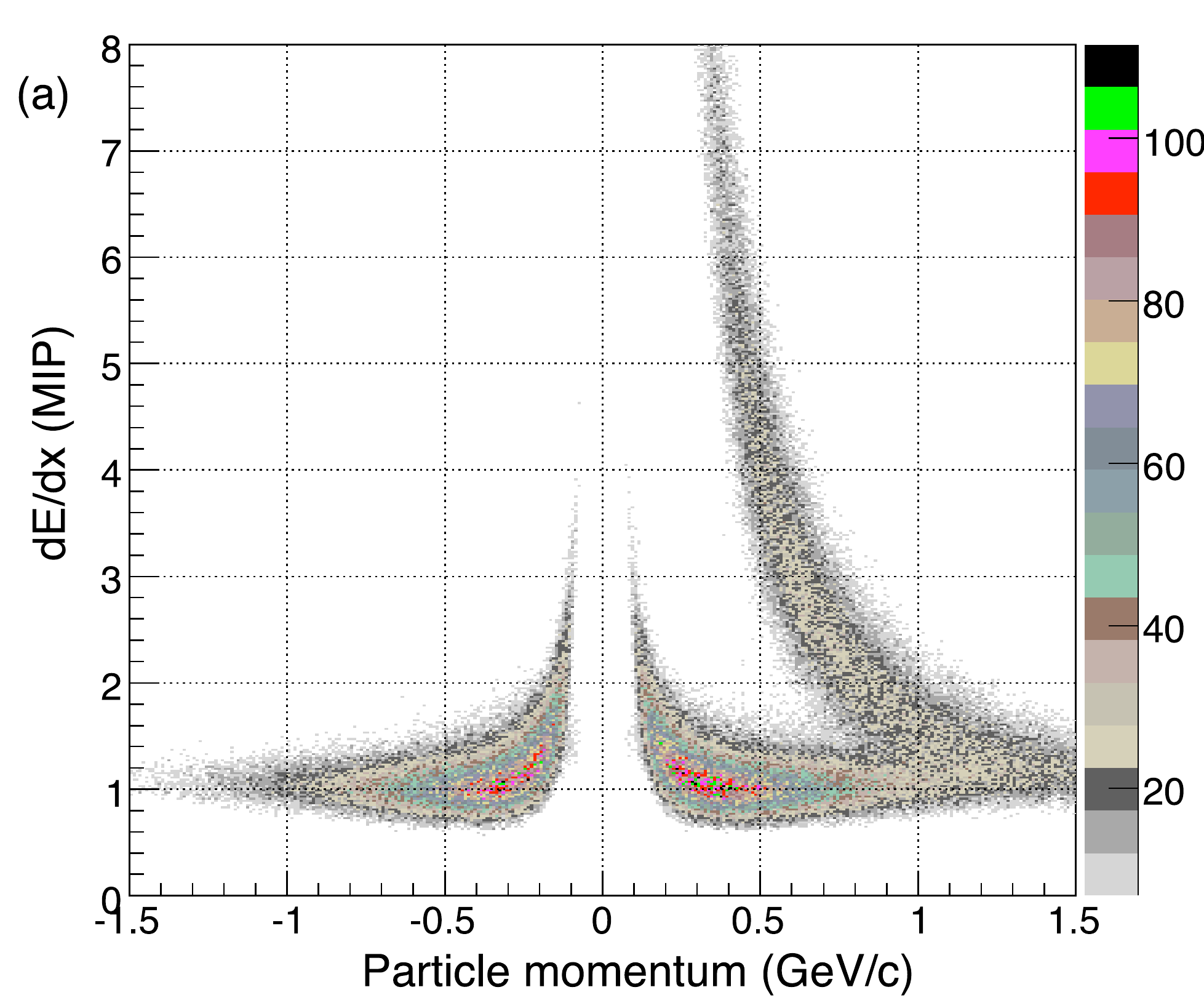} &
\includegraphics[width=7.8cm]{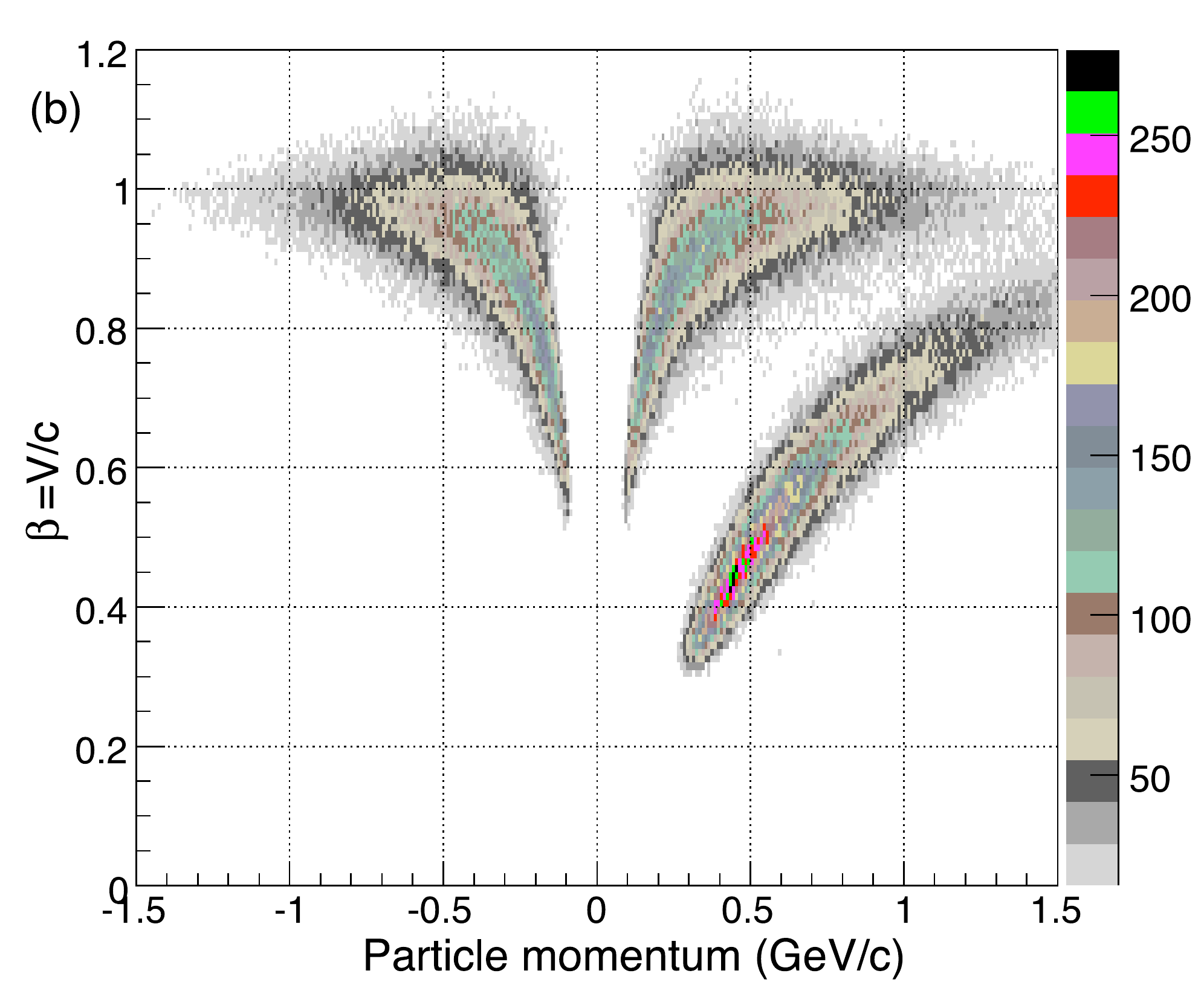} 
\end{tabular}
\end{center}
\caption{Specific ionization d$E$/d$x$ [in units of minimum-ionizing 
pulse height] (a) and velocity $\beta$ (b) versus momentum [GeV/{\it c}], 
for positive and negative tracks in $+8.9$~GeV/{\it c} data.}
\label{dedxandbeta}
\end{figure}

As for the separation of pions from protons: the 
resolution of \dedx\ in the TPC is 16\% for a track length
of 300~mm, and the system time-of-flight resolution is 175~ps.  
Figure~\ref{dedxandbeta} (a) shows the specific ionization
\dedx , measured by the TPC, and Fig.~\ref{dedxandbeta} (b) the 
relative velocity $\beta$ 
from the RPC time of flight, of positive and negative secondaries, 
as a function of the momentum measured in the TPC. The figures 
demonstrate that, in general, protons and pions are well separated.
They also underline the importance of the complementary 
separation by RPC time of flight at large particle momentum. 
The average values of \dedx\ and $\beta$ agree well with 
theoretical expectations, thus confirming the validity
of the absolute scales of momentum, \dedx , and time of flight.

\subsection{Acceptance and migration}
\label{acceptanceandmigration}

We discuss differences of inclusive spectra 
between data and simulated data in terms of the distribution
in the polar angle $\theta$, for different ranges of $p_{\rm T}$.
The primary reason for this choice is that disagreements show up  
most clearly in $\theta$. At the same time, $\theta$ is a 
well-measured experimental quantity. We also consider 
that the $\theta$ distribution provides the clue to the
origin of the disagreements. 

\begin{figure}[h]
\begin{center}
\begin{tabular}{cc}
\includegraphics[width=7.5cm]{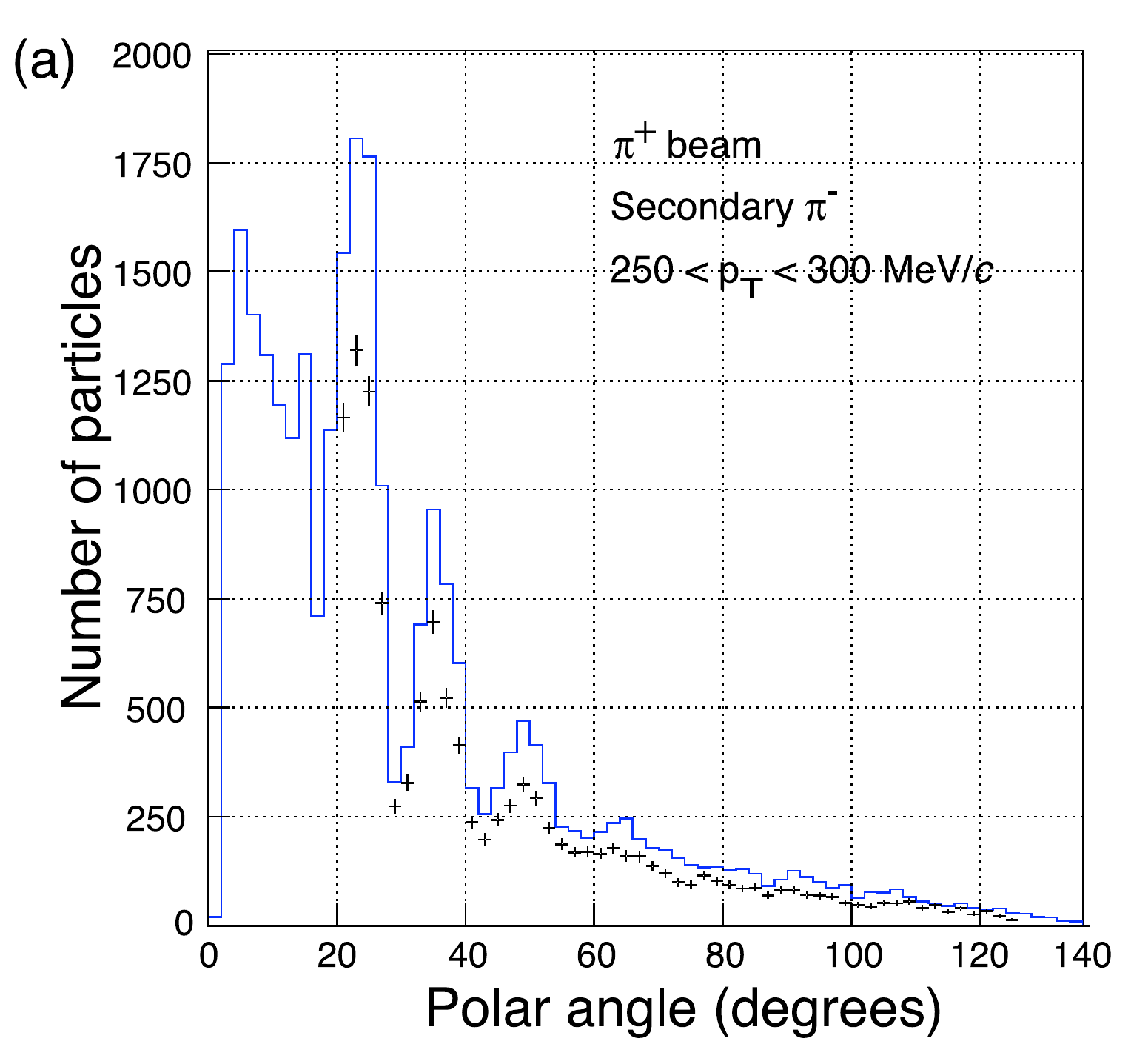} &
\includegraphics[width=7.5cm]{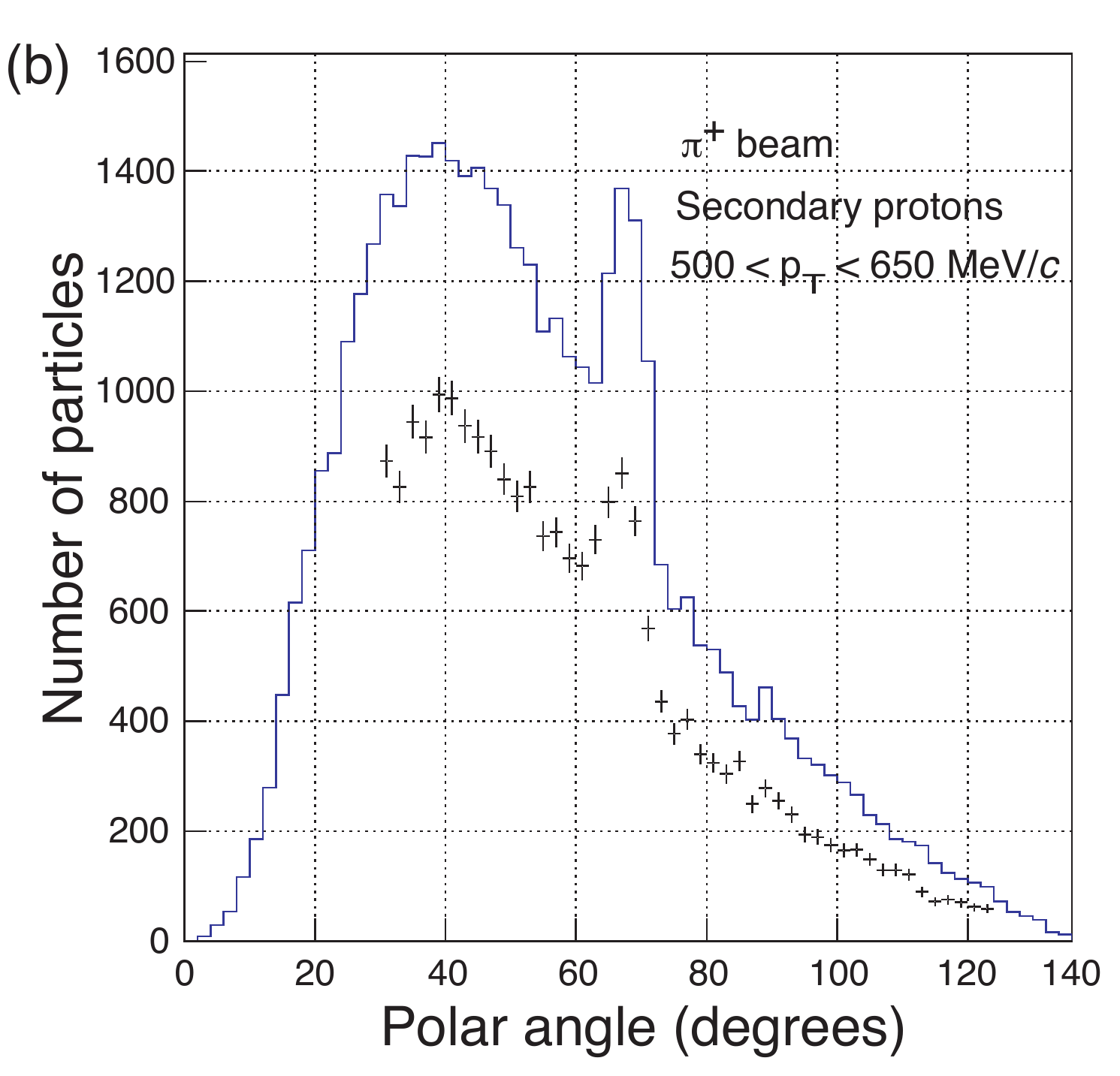} 
\end{tabular}
\end{center}
\caption{Comparison of Monte Carlo generated (full lines) with
Monte Carlo accepted (crosses) tracks; (a) polar-angle distributions of $\pi^-$ for
incoming $\pi^+$, and (b) of protons for incoming $\pi^+$.}
\label{generatedvsaccepted}
\end{figure}

The physics performance parameters in the transverse momentum
$p_{\rm T}$ and polar angle $\theta$ are so good that 
finite resolution, or a small dependence of acceptance cuts 
on $p_{\rm T}$ or $\theta$, does not 
appreciably affect the comparison of data with 
simulated data (the chosen ranges of $p_{\rm T}$ exceed by 
a factor of two or more the $p_{\rm T}$ resolution, and the
chosen bin size of 
$2^\circ$ ($35$~mrad) of $\theta$ exceeds
by a factor of two the $\theta$ resolution).    
That is substantiated in Fig.~\ref{generatedvsaccepted}
which shows examples of the dominant disagreements between
data and simulated data: patterns reminiscent of diffractive
scattering [Fig.~\ref{generatedvsaccepted} (a)] and of elastic 
scattering [Fig.~\ref{generatedvsaccepted} (b)]. The full lines
show the Monte Carlo generated polar-angle distributions, 
while the crosses show the same after acceptance cuts,
particle identification cuts, and
with resolution effects included. Obviously, the structures 
seen in the simulated data are not appreciably altered  
by experimental acceptance and resolution.

We conclude that for the
comparison of the shapes of inclusive particle spectra between
data and simulated data---which is the purpose of this paper---
it is sufficient to compare data with Monte Carlo generated 
distributions, without correction of 
losses from acceptance cuts and of migration stemming from 
finite resolution. Also, the comparison is intentionally restricted to kinematical regions 
where there is ample and unambiguous separation of pions from protons\footnote{Absolutely
normalized double-differential cross-sections, obtained after
due corrections for acceptance and migration, and making use 
of proper weights for particle identification and therefore
spanning larger ranges of kinematical parameters than
discussed in this paper, will be presented in forthcoming 
papers.}. To identify a secondary particle as a pion or as a proton
it is required that the measured d$E$/d$x$ and time of flight
are both consistent with the given particle hypothesis.
In addition, the time of flight has to be inconsistent
with the opposite hypothesis. For particles without
either d$E$/d$x$ or time-of-flight measurement, the cut on 
the available variable is tightened. Within the accepted phase 
space, the particle identification efficiency is between 70\% and 90\%, while the contribution
from wrong particle identification is below 5\%.

\section{Data versus simulation from selected Geant4 physics lists}

The combination of the choice of hadron generators with the 
choice of incoming beam particles and the choice of secondary hadrons,
leads to a large a number of possible plots. With a view to simplifying
matters, we select for several physics lists two plots each that are 
representative for
the agreement and disagreement, respectively, between 
data and simulation.

\begin{figure}[h]
\begin{center}
\begin{tabular}{cc} 
\includegraphics[width=7.6cm]{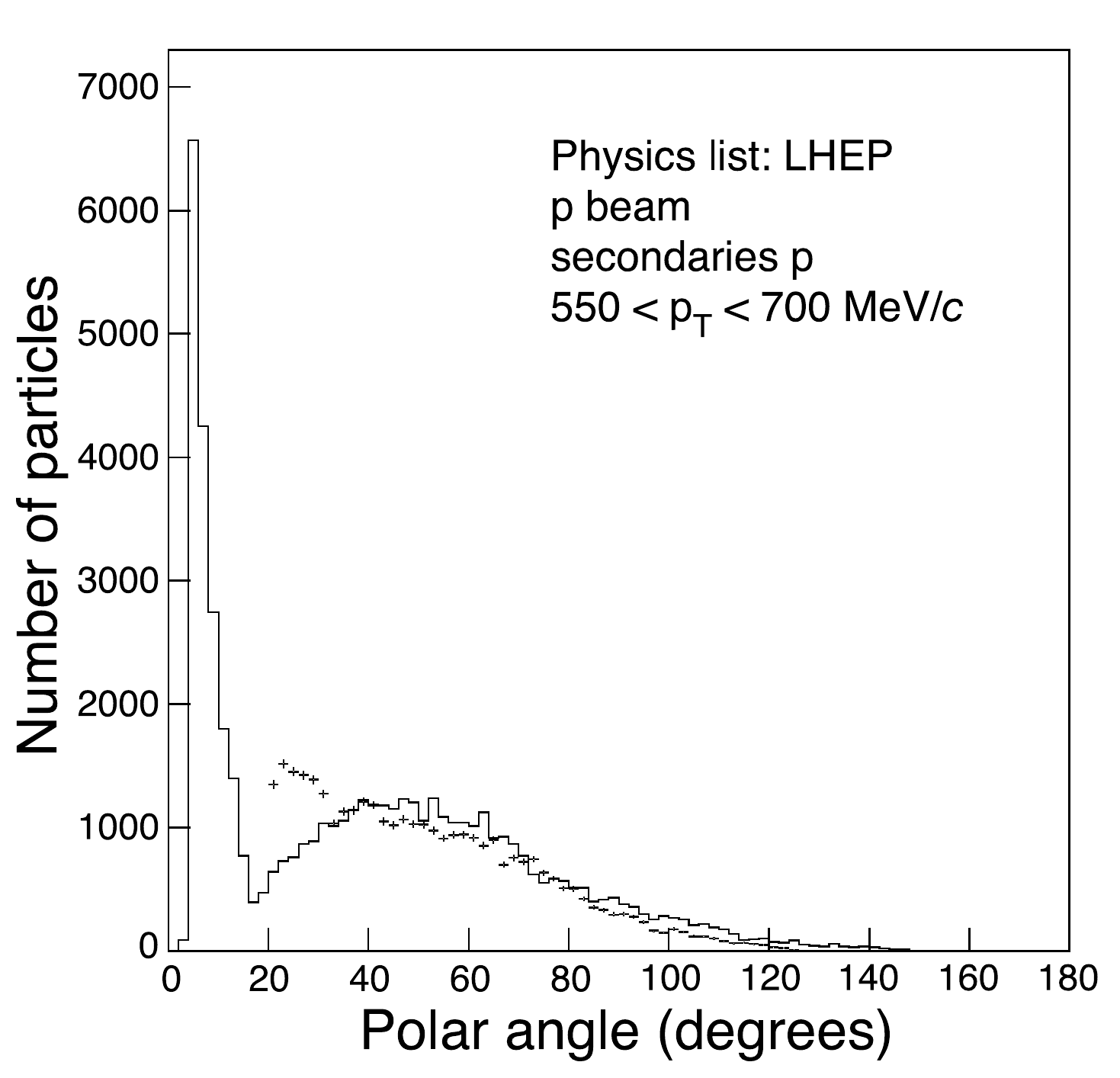} &
\includegraphics[width=7.6cm]{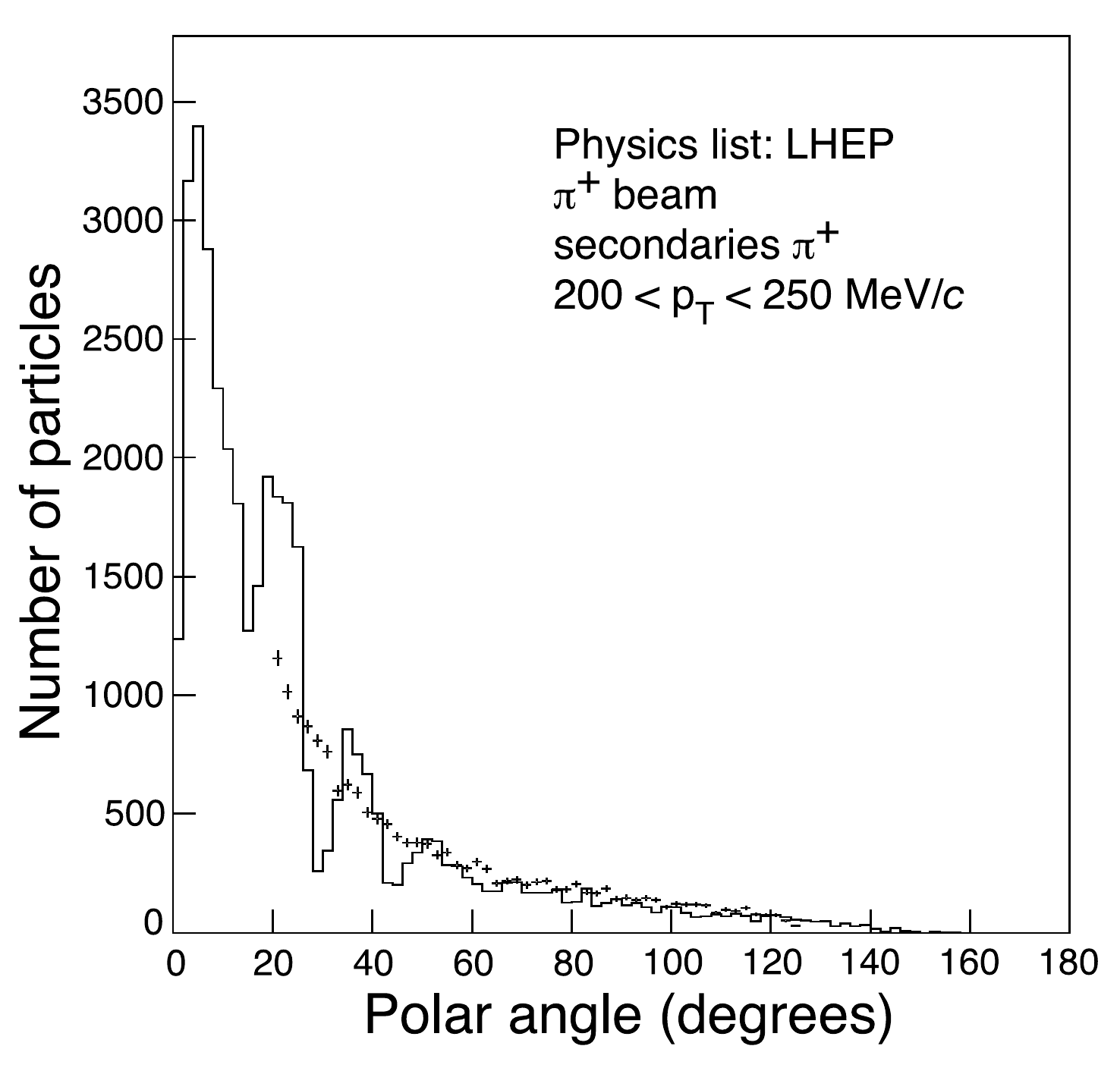} 
\end{tabular}
\end{center}
\caption{LHEP physics list; polar-angle distributions of protons for
incoming protons (left panel), and of $\pi^+$ for incoming $\pi^+$
(right panel).}
\label{complhep}
\end{figure}

\begin{figure}[h]
\begin{center}
\begin{tabular}{cc} 
\includegraphics[width=7.5cm]{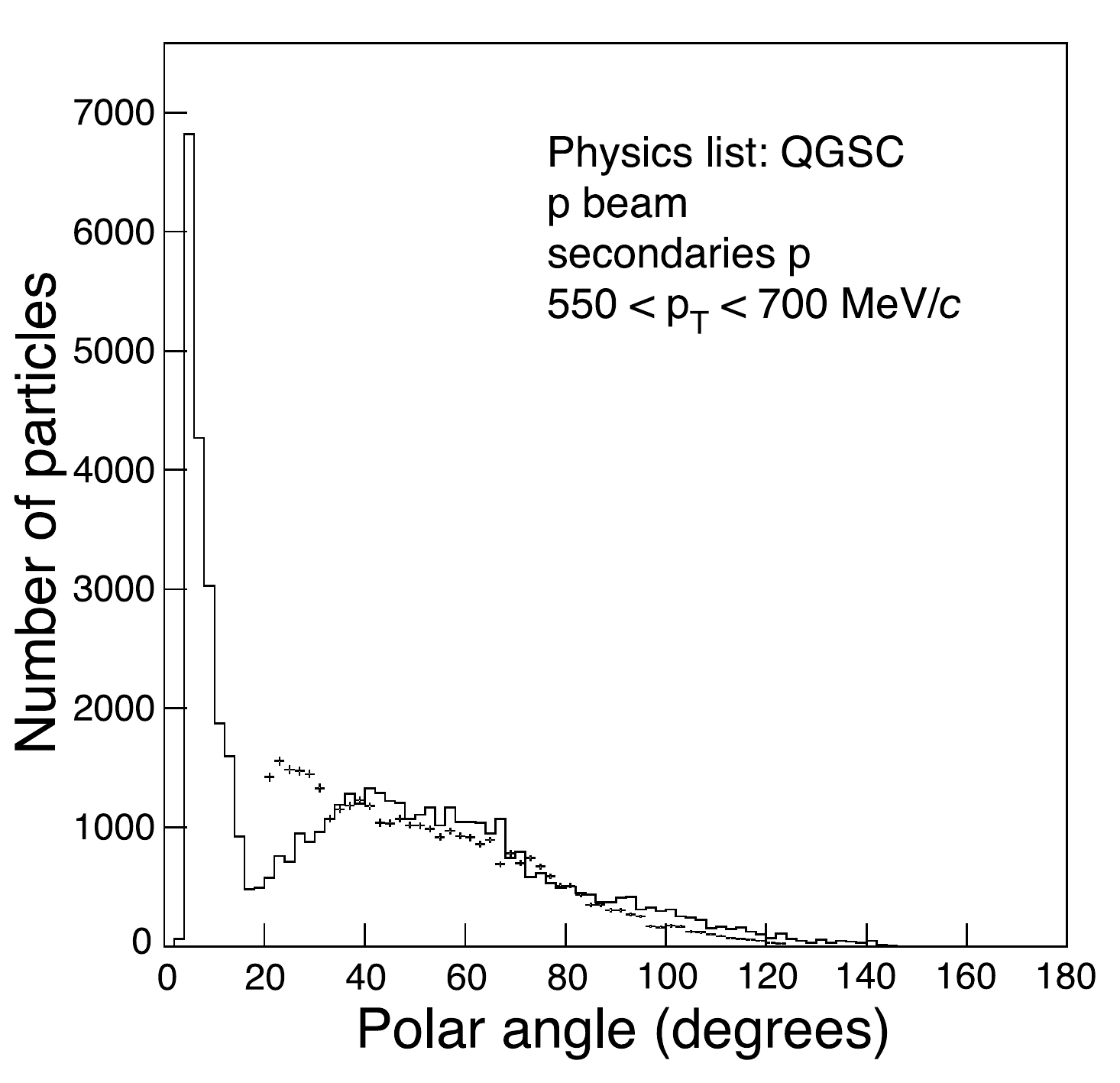} &
\includegraphics[width=7.6cm]{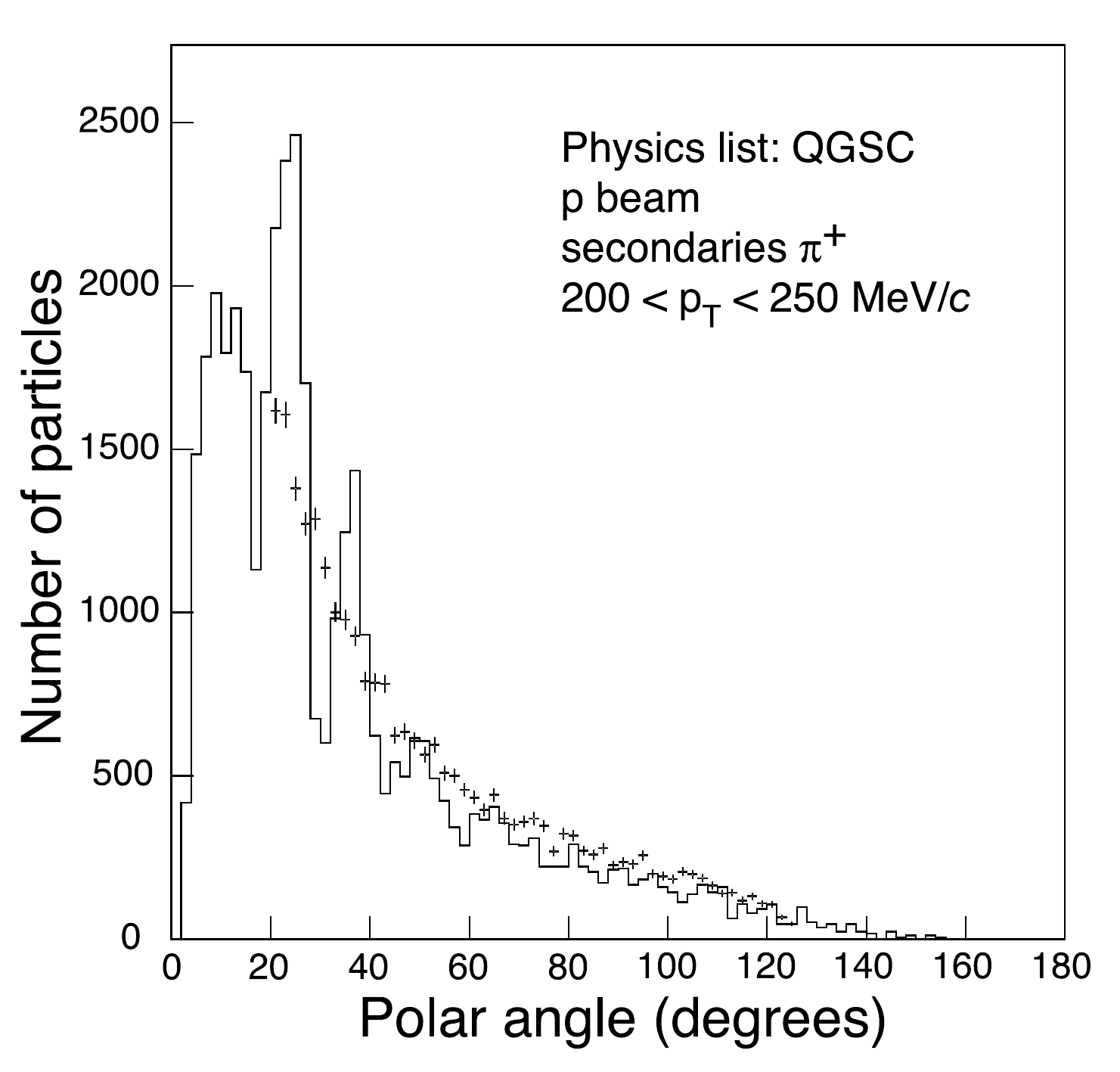} 
\end{tabular}
\end{center}
\caption{QGSC physics list; polar-angle distributions of protons for
incoming protons (left panel), and of $\pi^+$ for incoming protons
(right panel).}
\label{compqgsc}
\end{figure}

\begin{figure}[h]
\begin{center}
\begin{tabular}{cc}
\includegraphics[width=7.5cm]{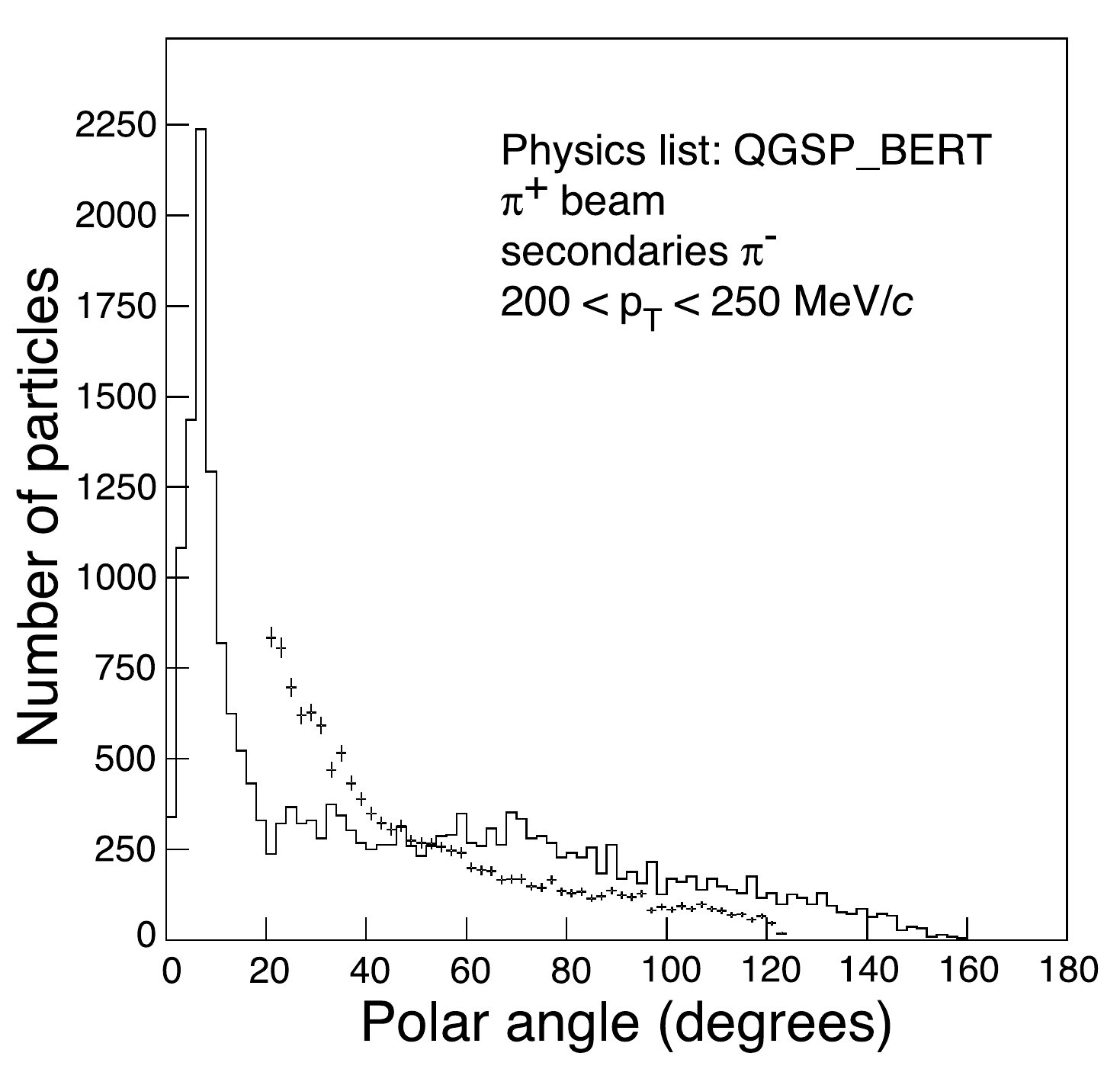} &
\includegraphics[width=7.5cm]{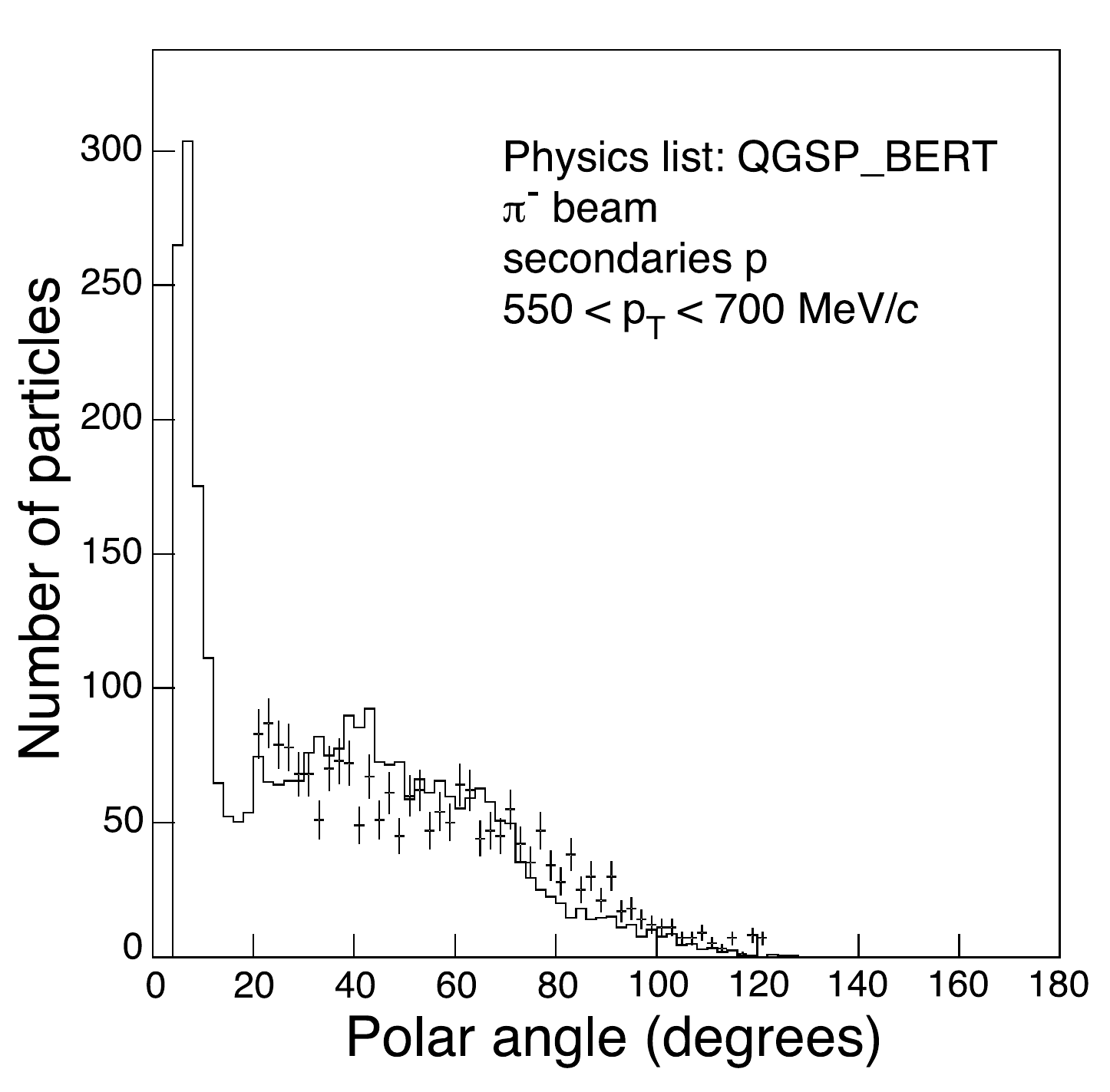} 
\end{tabular}
\end{center}
\caption{QGSP\_BERT physics list; polar-angle distributions of $\pi^-$
for incoming $\pi^+$ (left panel), and of protons for incoming $\pi^-$
(right panel).}
\label{compqgsp_bert}
\end{figure}

\begin{figure}[h]
\begin{center}
\begin{tabular}{cc} 
\includegraphics[width=7.5cm]{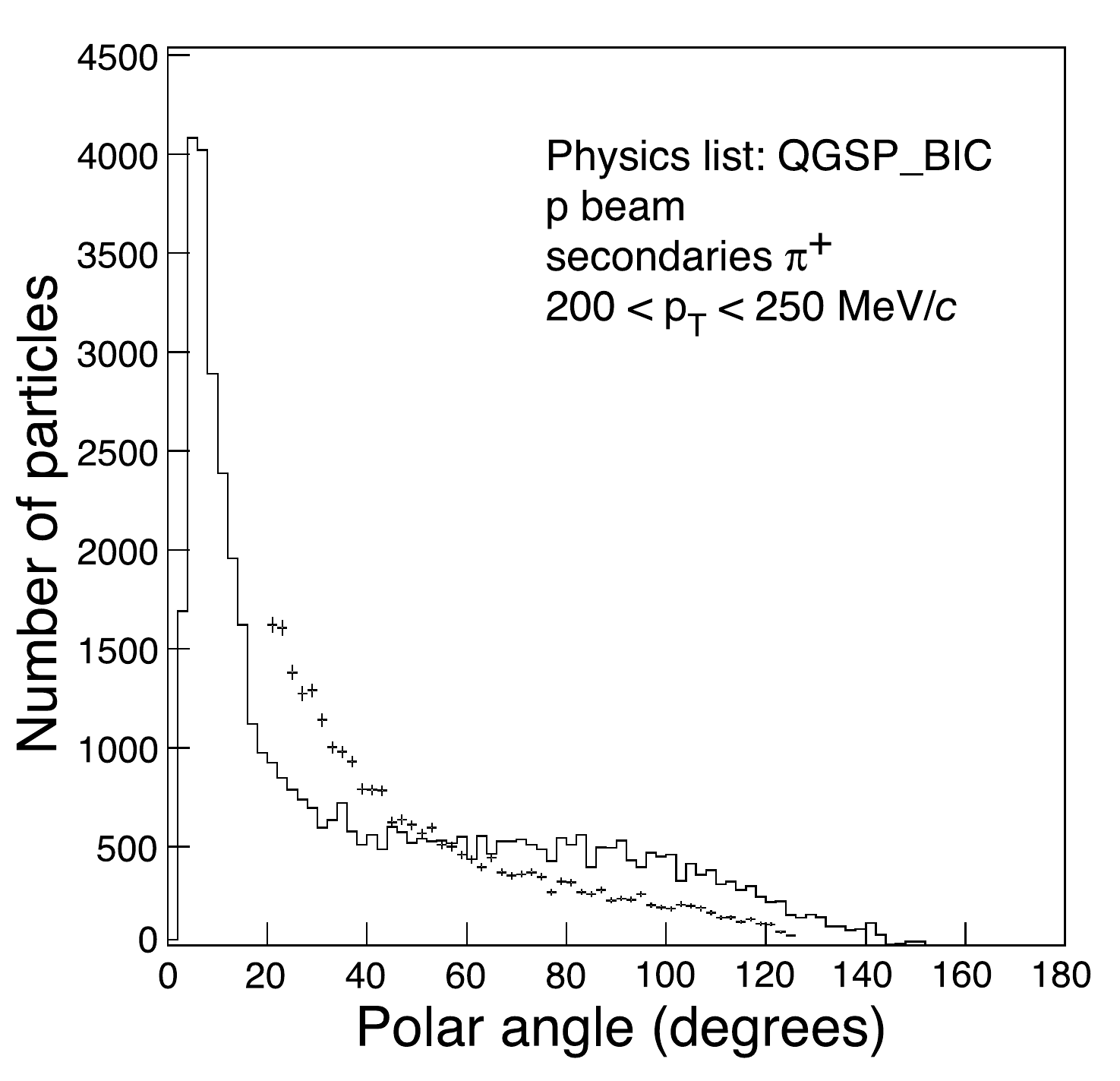} &
\includegraphics[width=7.5cm]{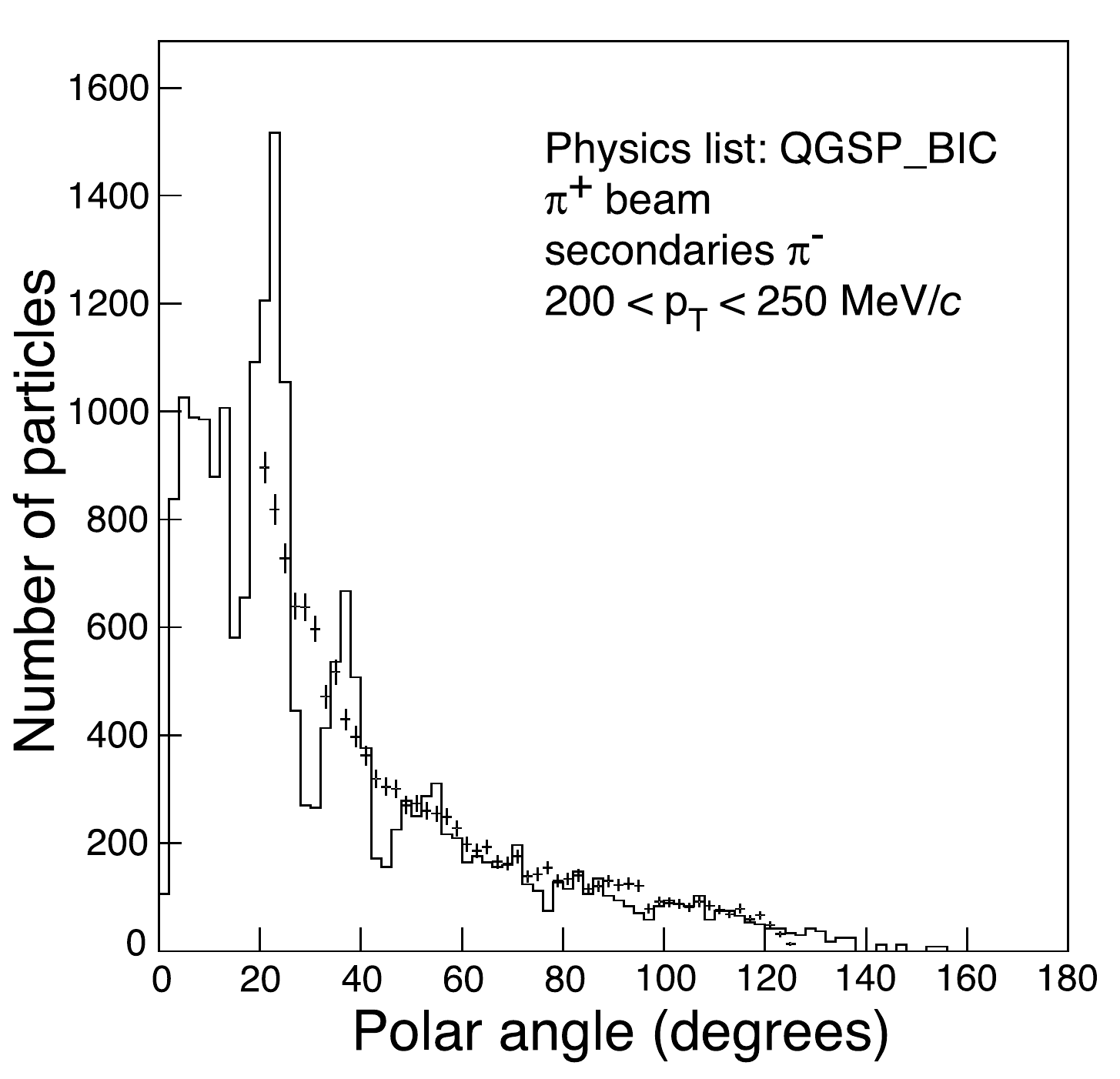} 
\end{tabular}
\end{center}
\caption{QGSP\_BIC physics list; polar-angle distributions of $\pi^+$
for incoming protons (left panel), and of $\pi^-$ for incoming $\pi^+$
(right panel).}
\label{compqgsp_bic}
\end{figure}

\begin{figure}[h]
\begin{center}
\begin{tabular}{cc} 
\includegraphics[width=7.5cm]{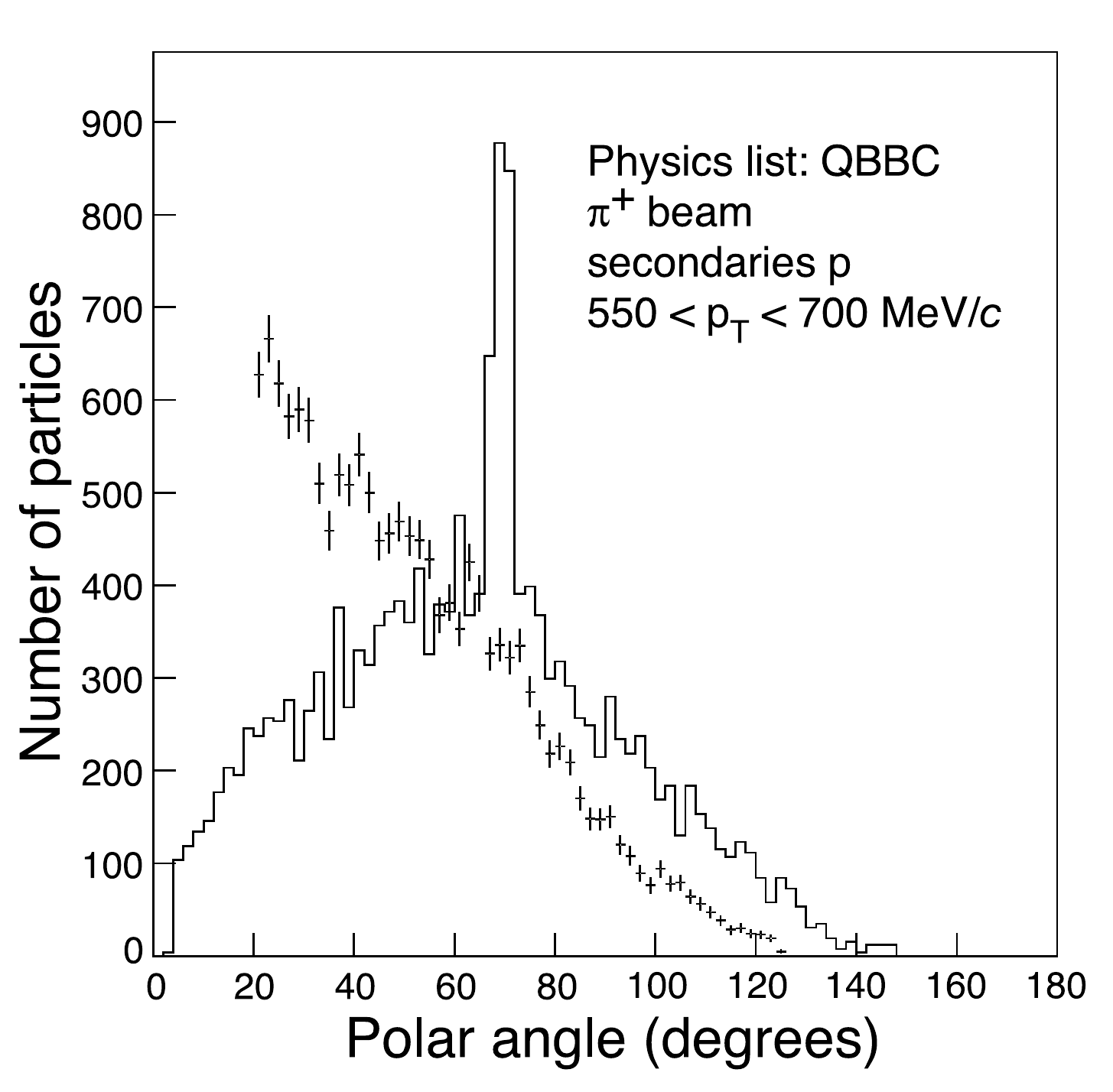} &
\includegraphics[width=7.5cm]{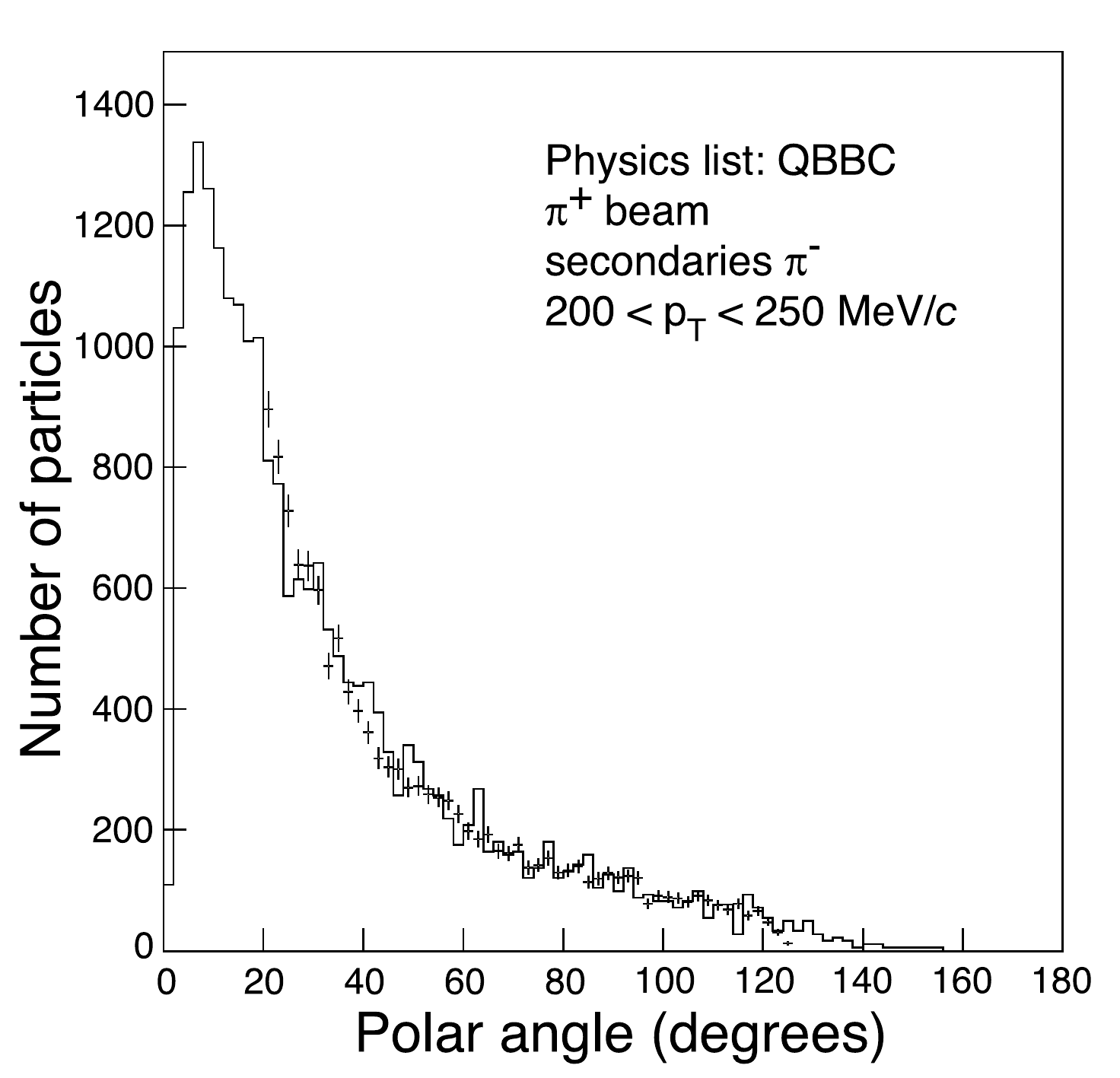} 
\end{tabular}
\end{center}
\caption{QBBC physics list; polar-angle distributions of protons
for incoming $\pi^+$ (left panel), and of $\pi^-$ for incoming $\pi^+$
(right panel).}
\label{compqbbc}
\end{figure}

\begin{figure}[h]
\begin{center}
\begin{tabular}{cc} 
\includegraphics[width=7.5cm]{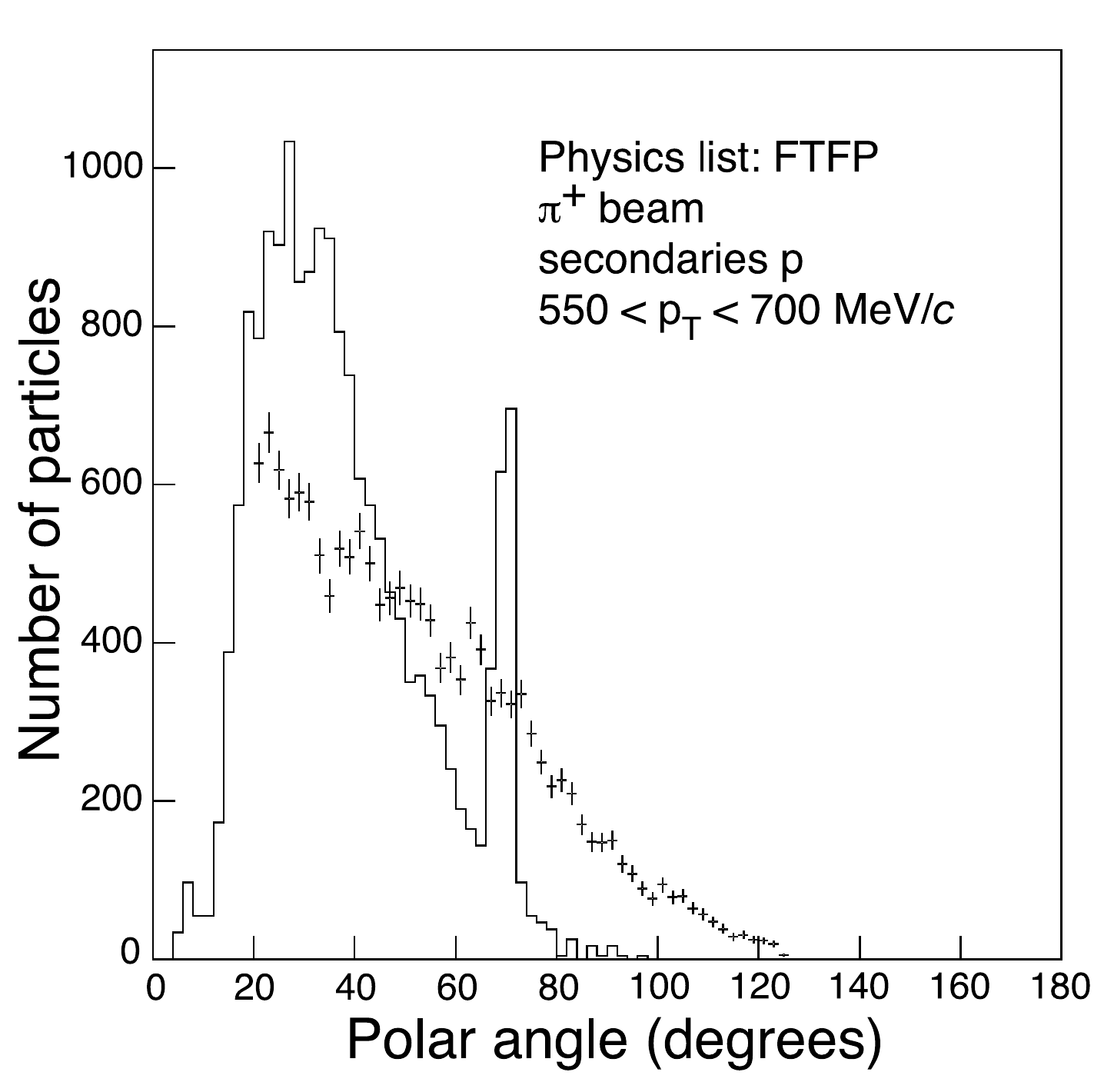} &
\includegraphics[width=7.5cm]{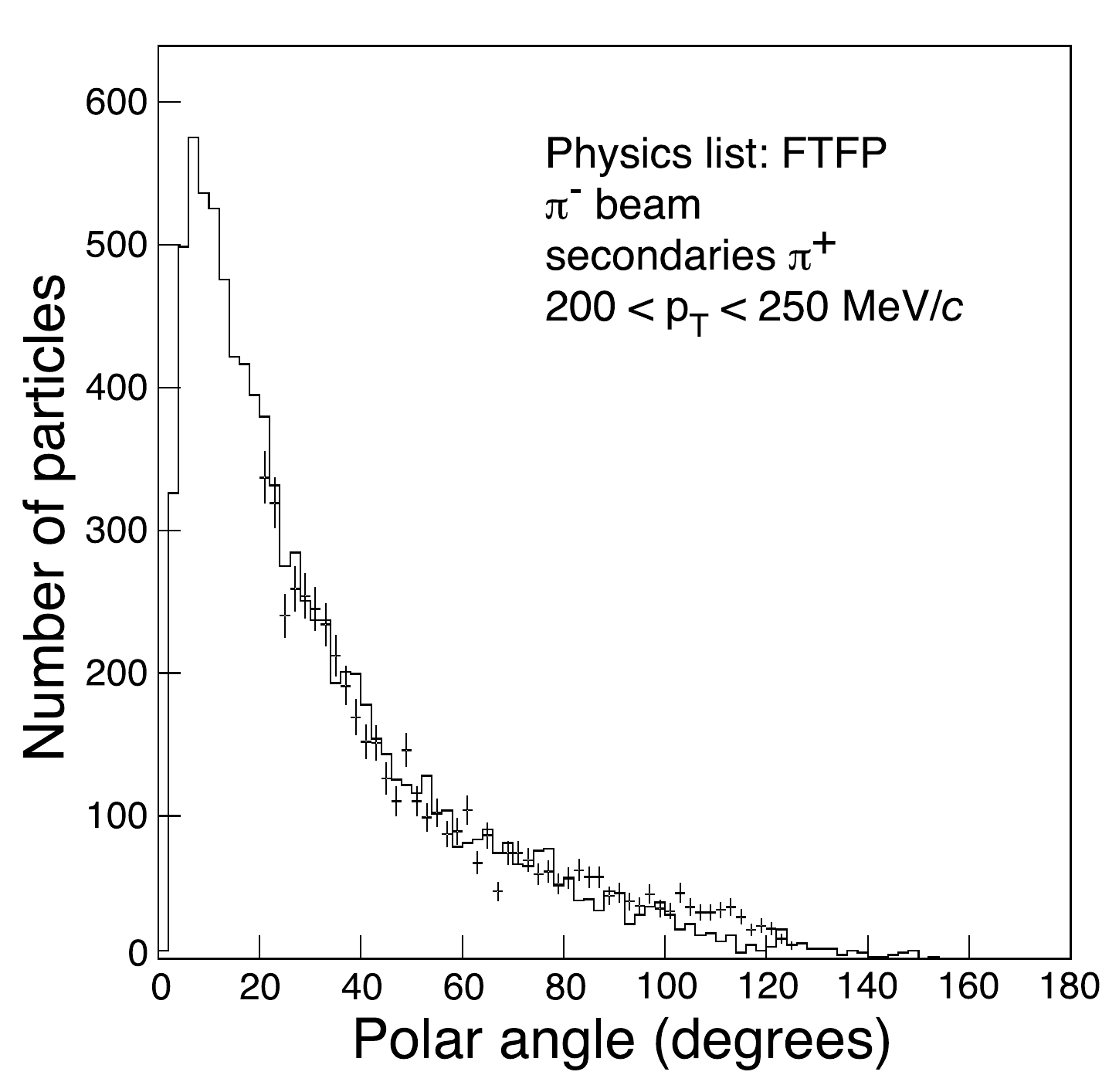} 
\end{tabular}
\end{center}
\caption{FTFP physics list; polar-angle distributions of protons
for incoming $\pi^+$ (left panel), and of $\pi^+$ for incoming $\pi^-$
(right panel).}
\label{compftfp}
\end{figure}

In all plots, we compare the Monte Carlo-generated $\theta$
distribution with the $\theta$ distribution of data. Positive
beam particles have $+8.9$~GeV/{\it c} momentum, and negative
beam particles have $-8.0$~GeV/{\it c} momentum. The target
is a 5\% $\lambda_{\rm abs}$ thick stationary beryllium target.
In Figs.~\ref{complhep} to \ref{compftfp}, data are shown as crosses
while simulated data are shown as full lines.
As justified in Section~\ref{acceptanceandmigration}, 
the data are not corrected for losses from acceptance cuts and
migration stemming from finite resolution. With a view to
emphasizing shape differences, the data are normalized to
the simulated data in the angular range 
$20^\circ < \theta < 125^\circ$.

Figures~\ref{complhep} to \ref{compftfp} show the comparisons
for several physics lists from Table~\ref{physicslistsoverview}: LHEP, 
QGSC, QGSP\_BERT, QGSP\_BIC,
QBBC, and FTFP.  

There are three distinct problems visible in the comparison of
$\theta$ distributions of data and simulated data: 
\begin{itemize}
\item an unphysical peak for
secondary protons near $\theta = 70^\circ$;
\item an unphysical diffraction-like pattern for secondary pions;
\item a poor agreement in the shape. 
\end{itemize}

The different physics lists behave differently with
respect to the type of disagreement, yet the unphysical peak for
secondary protons near $\theta = 70^\circ$, and the unphysical 
diffraction-like pattern for secondary pions appear as
dominant problems.   

\clearpage

With a view to elucidating the physics origin of these dominating problems, we examine more closely for the LHEP physics list
the disagreements between data and simulated data for different combinations of incoming and secondary particle. 

\begin{figure}[h]
\begin{center}
\begin{tabular}{ccc} 
\includegraphics[width=5cm]{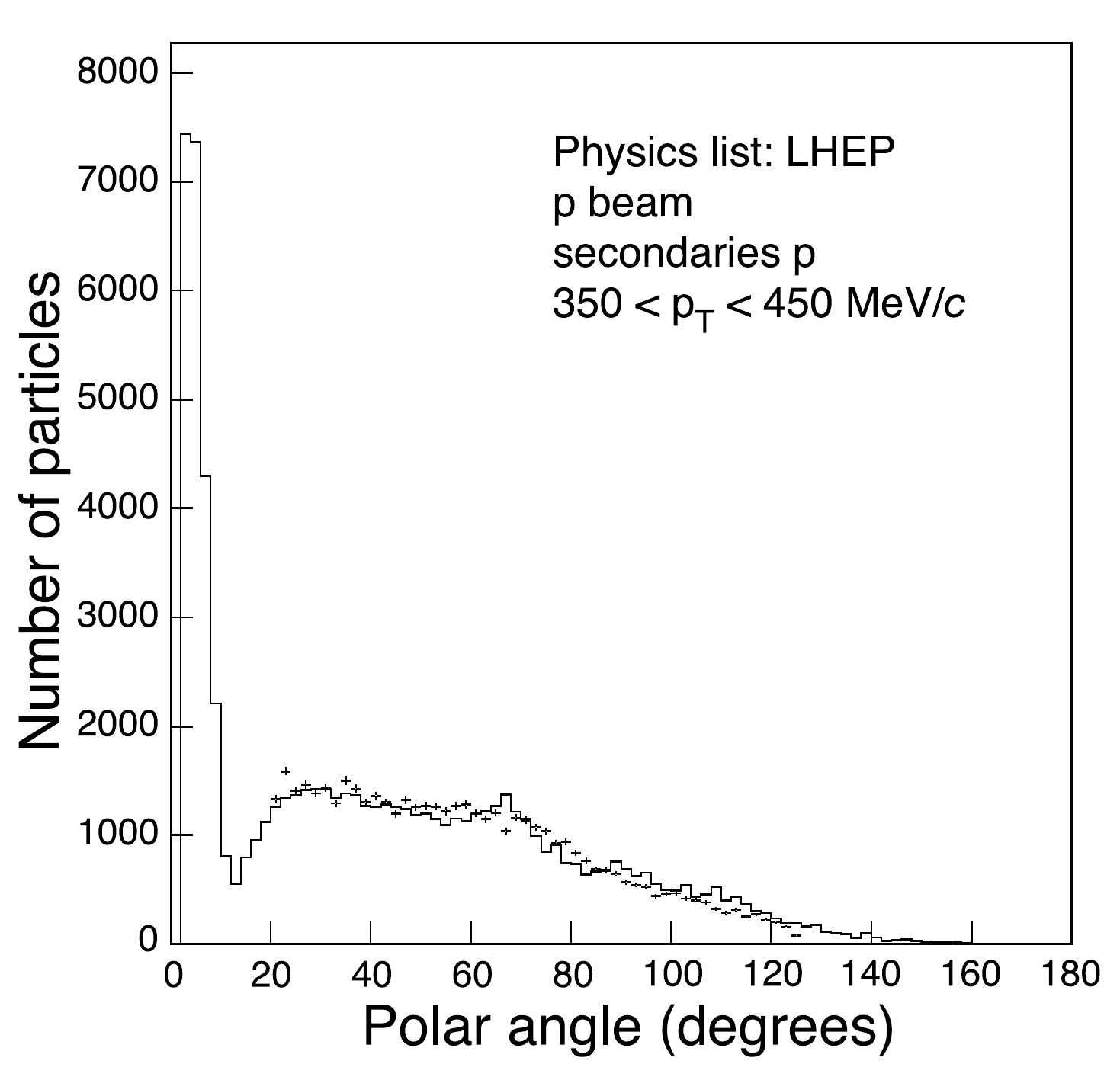} &
\includegraphics[width=5cm]{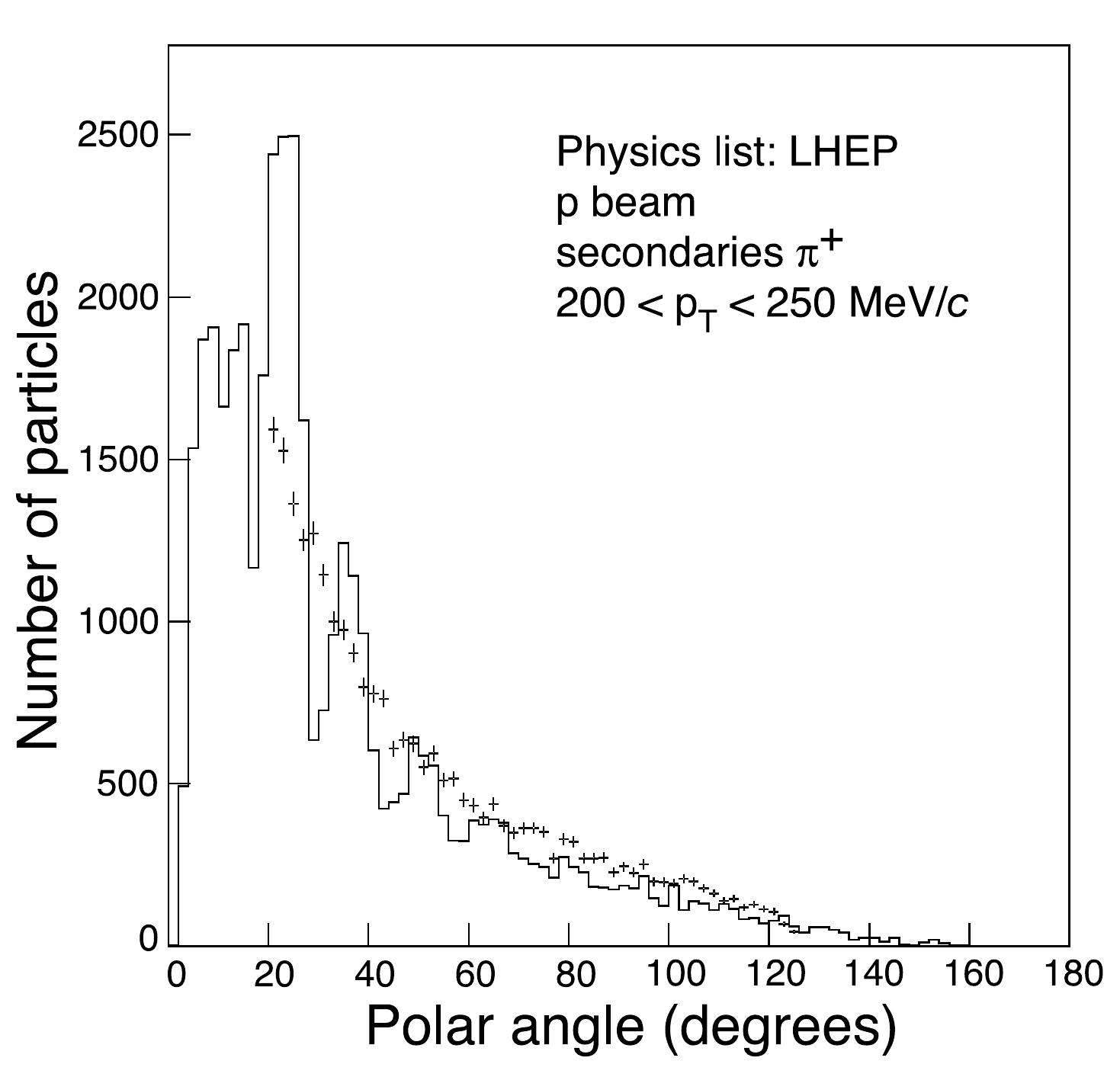} &
\includegraphics[width=5cm]{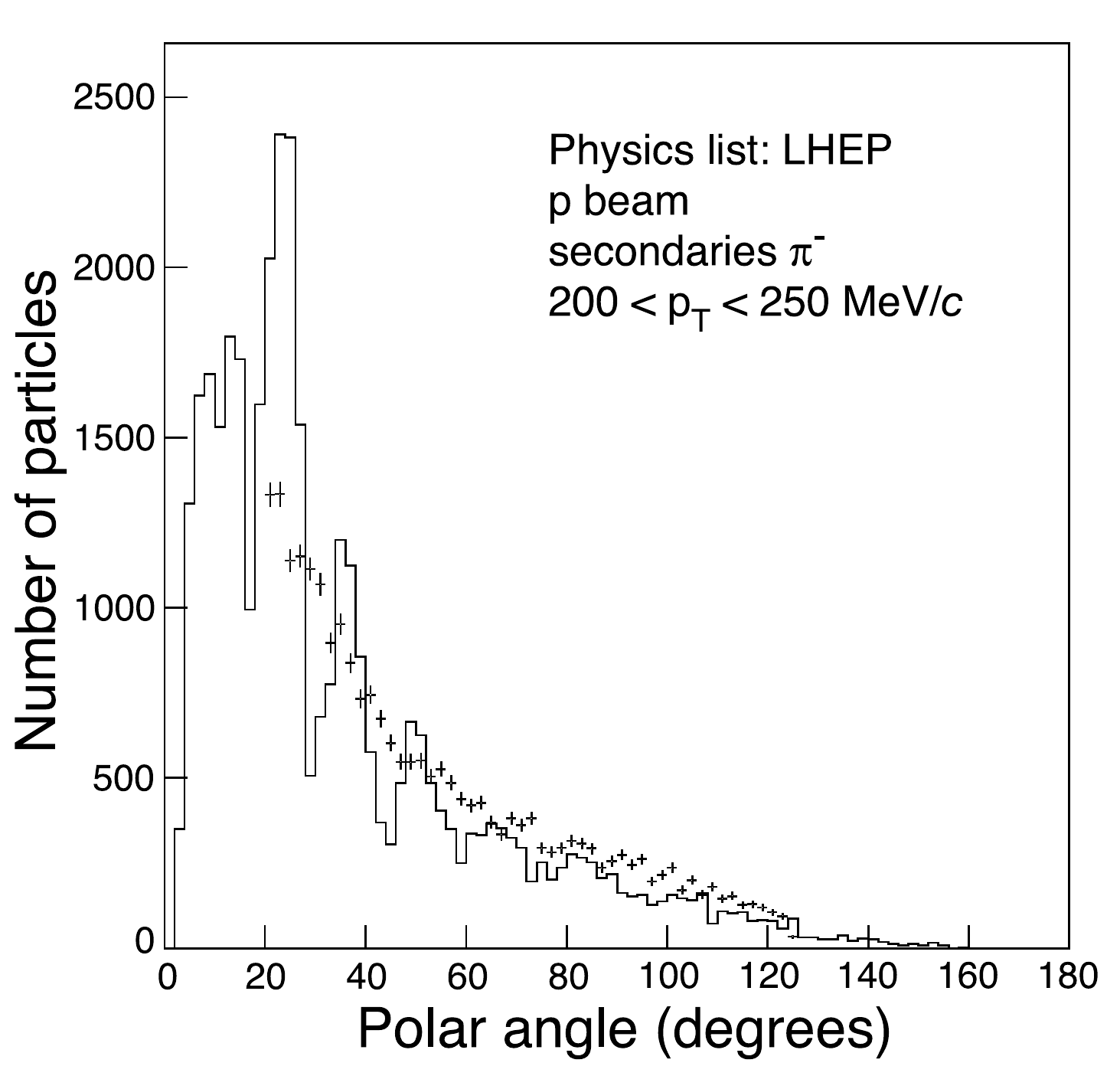} \\
\includegraphics[width=5cm]{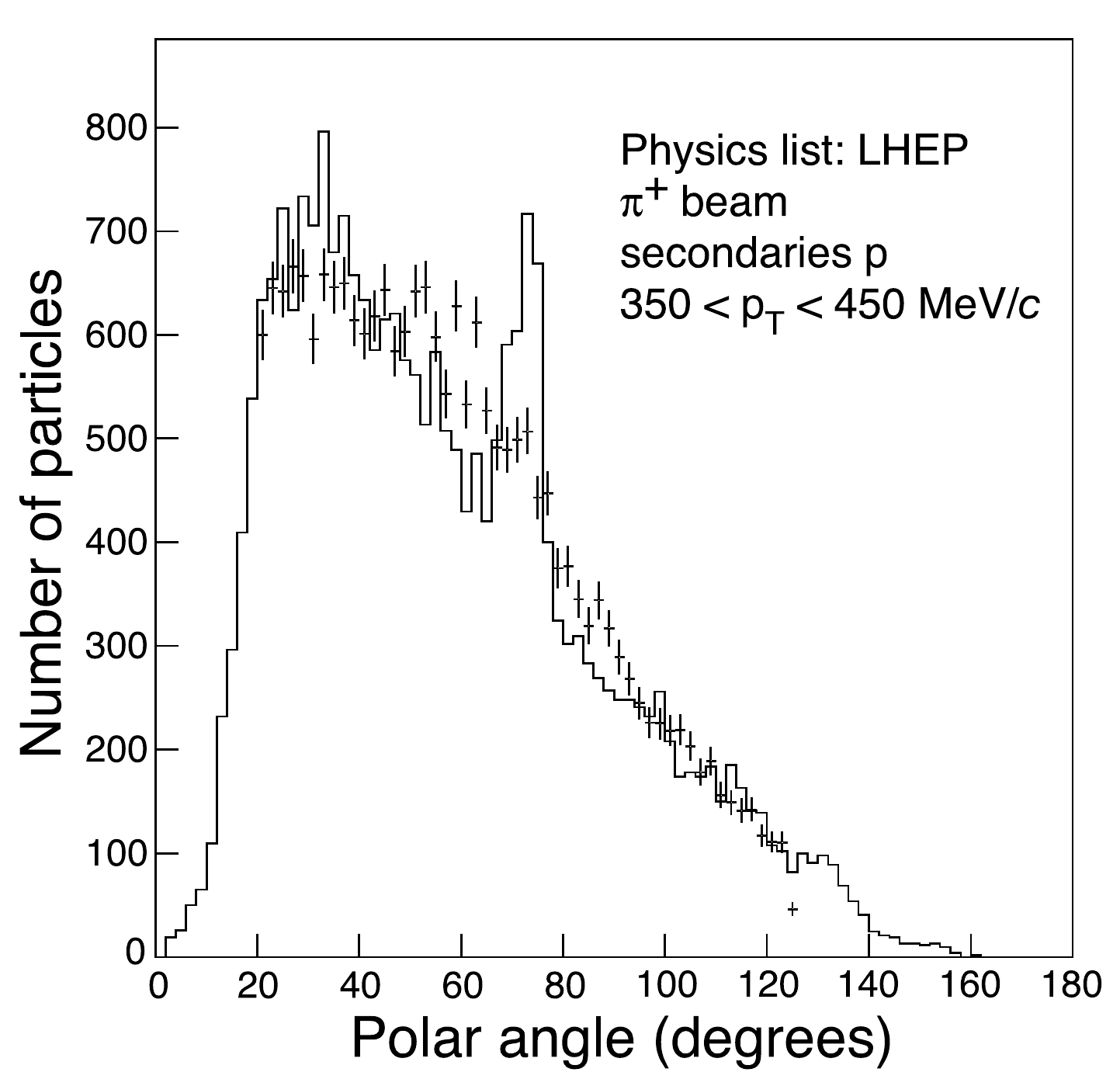} &
\includegraphics[width=5cm]{pi+_lhep_pi+_200to250.pdf} &
\includegraphics[width=5cm]{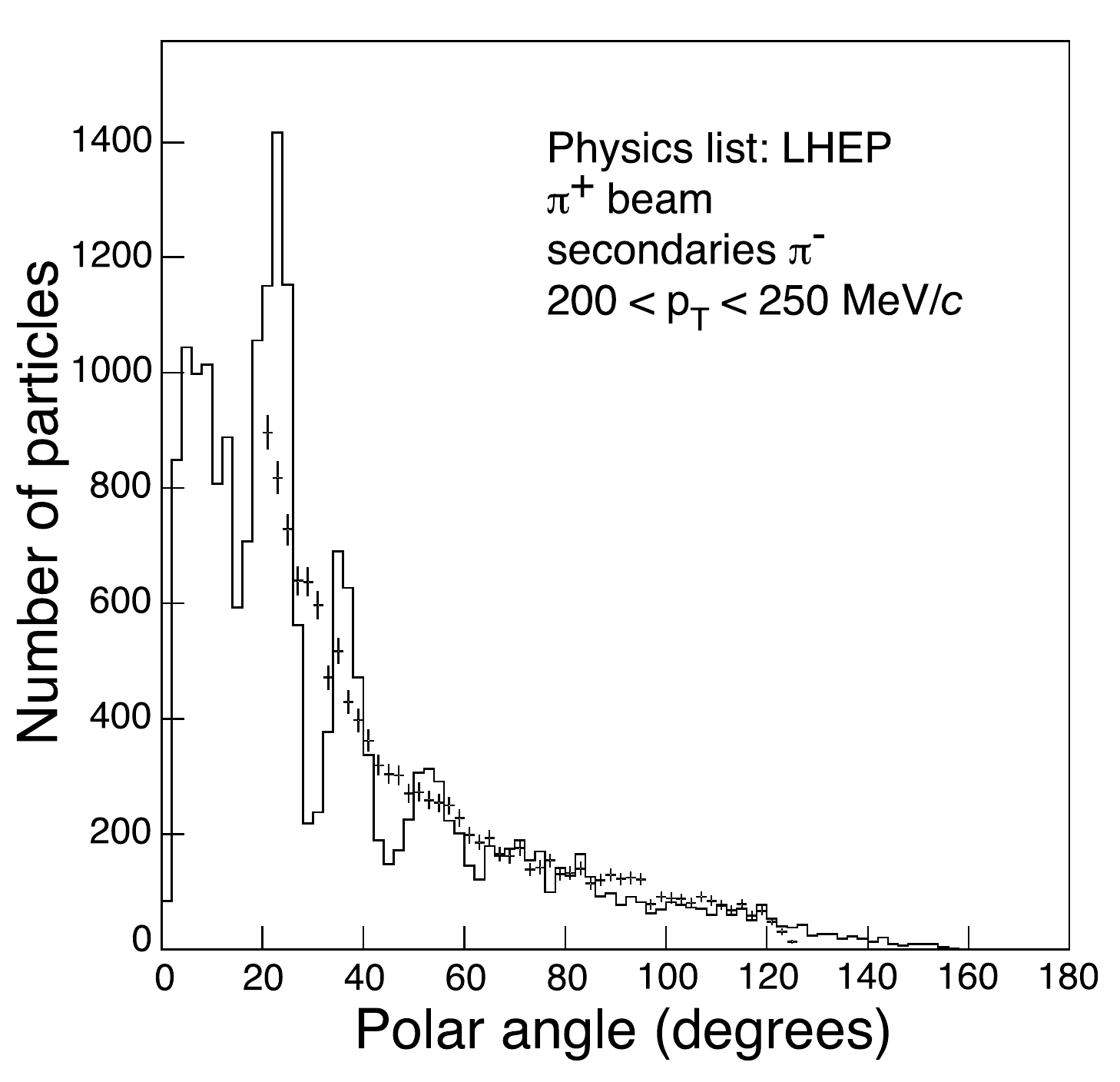} \\
\includegraphics[width=5cm]{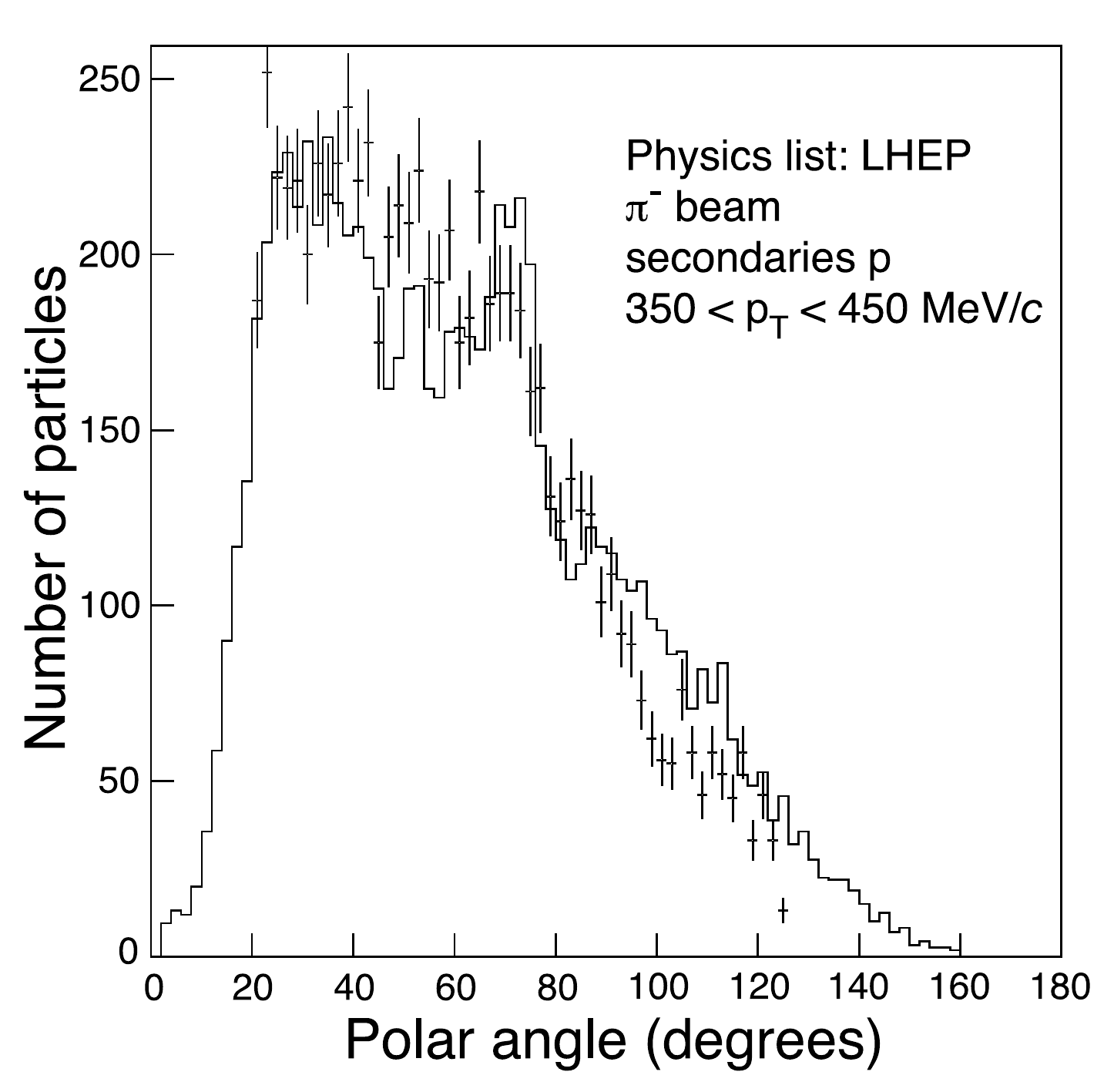} &
\includegraphics[width=5cm]{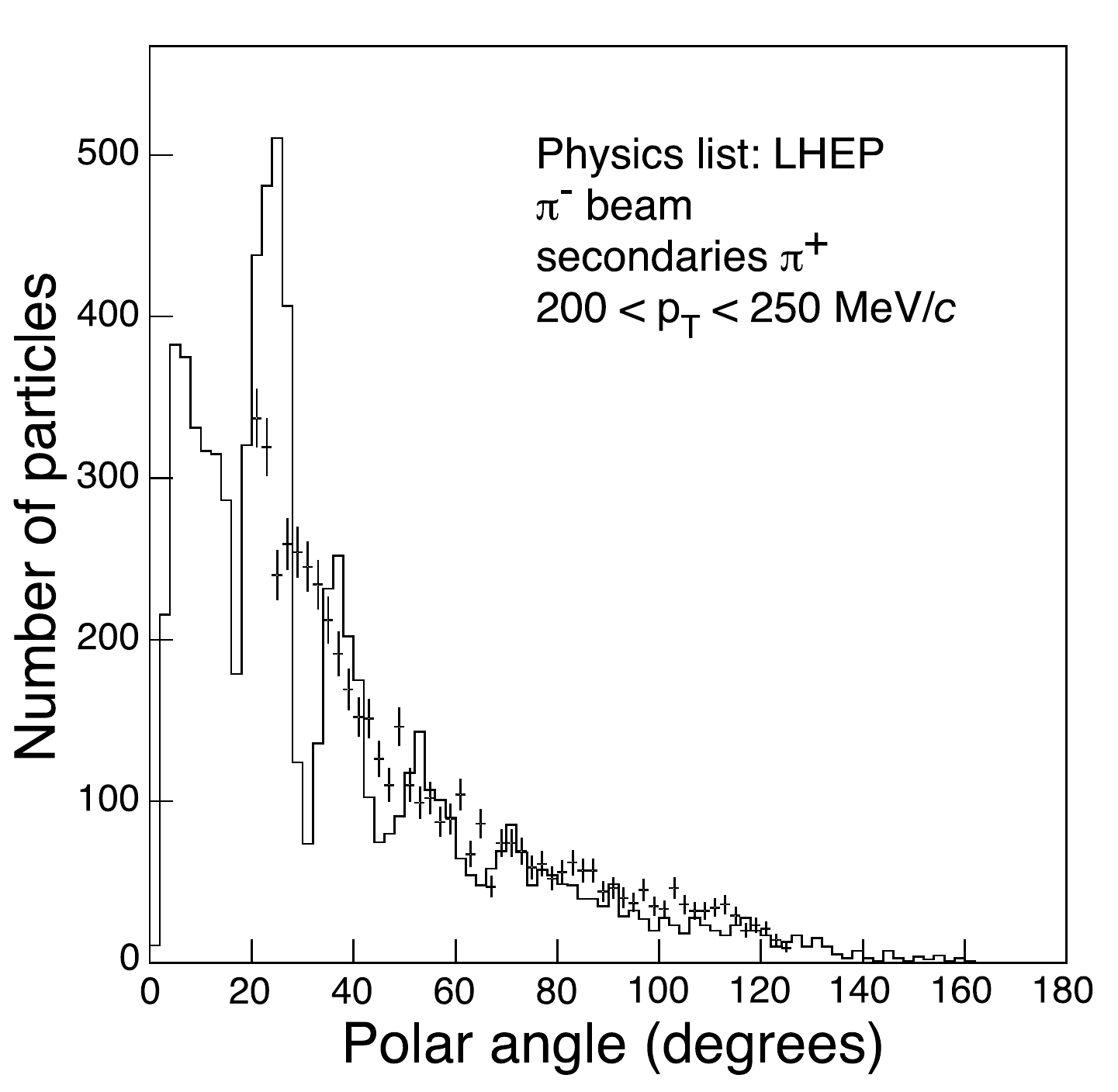} &
\includegraphics[width=5cm]{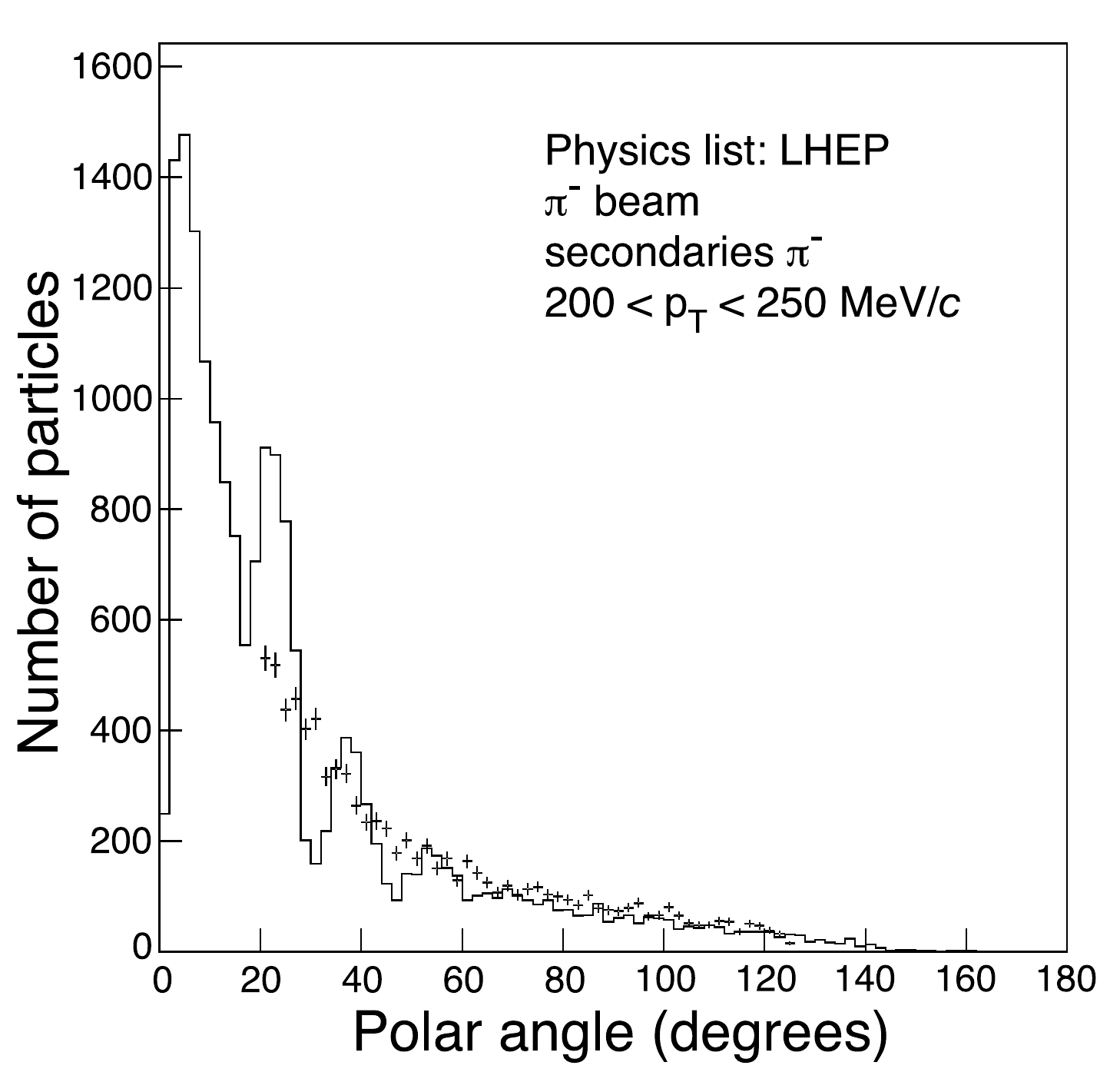} \\
\end{tabular}
\end{center}
\caption{LHEP physics list; polar-angle distributions of protons (left panels), $\pi^+$ (middle panels), and $\pi^-$ (right panels), 
for incoming protons (top row), incoming $\pi^+$ (middle row), and incoming $\pi^-$ (bottom row).}
\label{lhepbeamsecondary}
\end{figure}

Figure~\ref{lhepbeamsecondary} shows for a specific range of $p_{\rm T}$ all combinations of incoming and secondary protons, $\pi^+$ and $\pi^-$. 

\begin{figure}[h]
\begin{center}
\begin{tabular}{cc} 
\includegraphics[width=7.5cm]{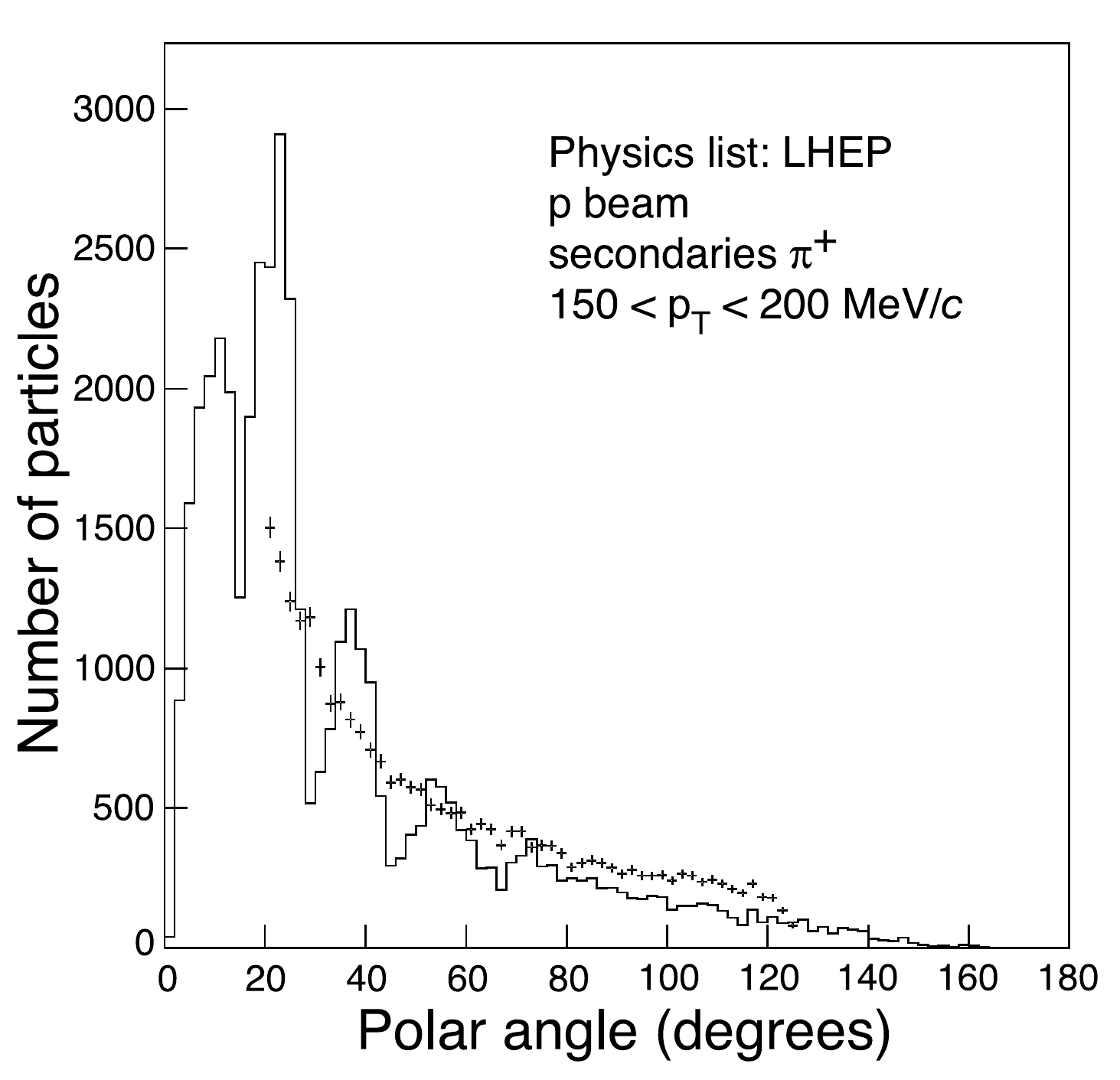} &
\includegraphics[width=7.5cm]{p_lhep_pi+_200to250.pdf} \\
\includegraphics[width=7.5cm]{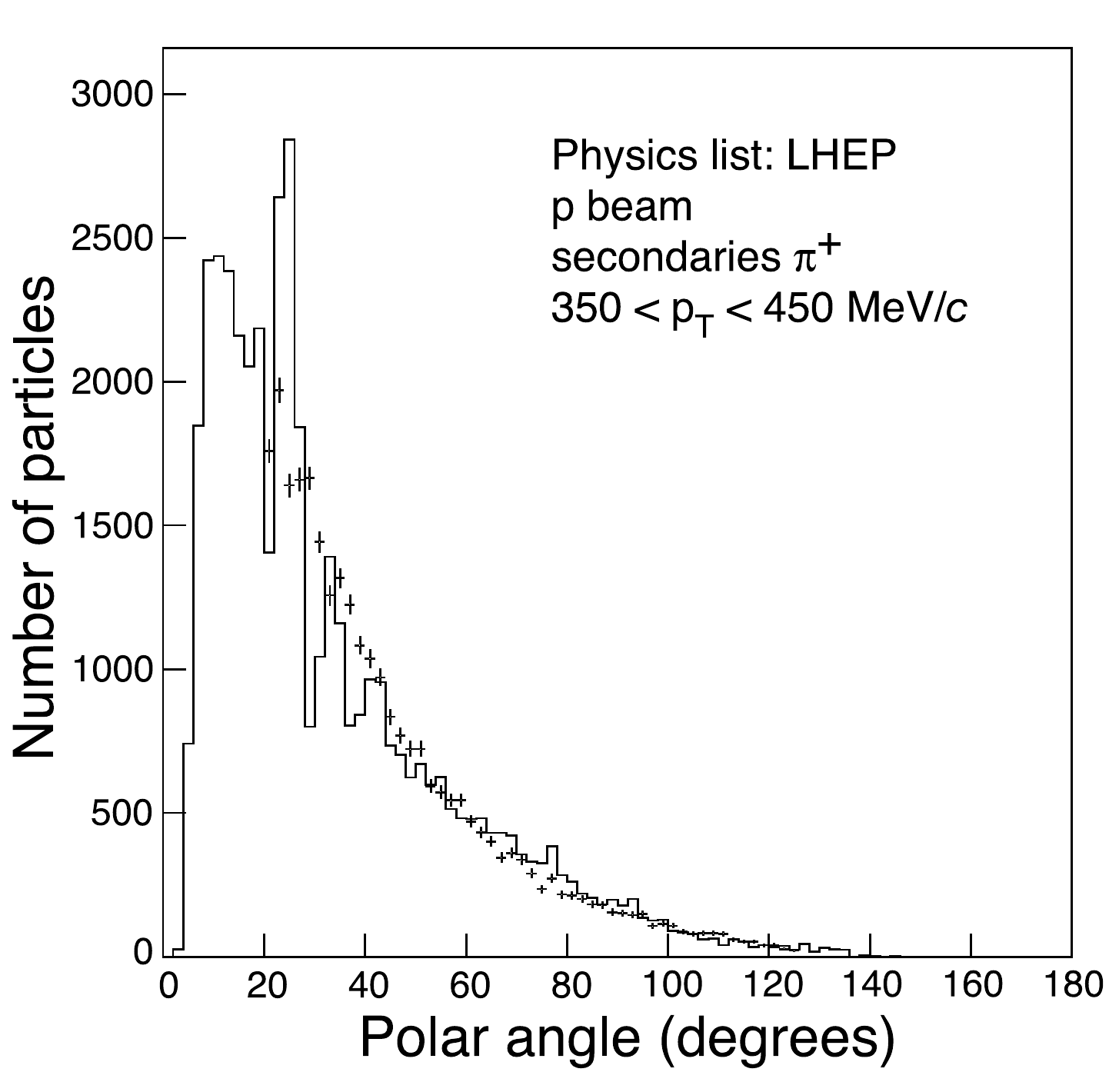} &
\includegraphics[width=7.5cm]{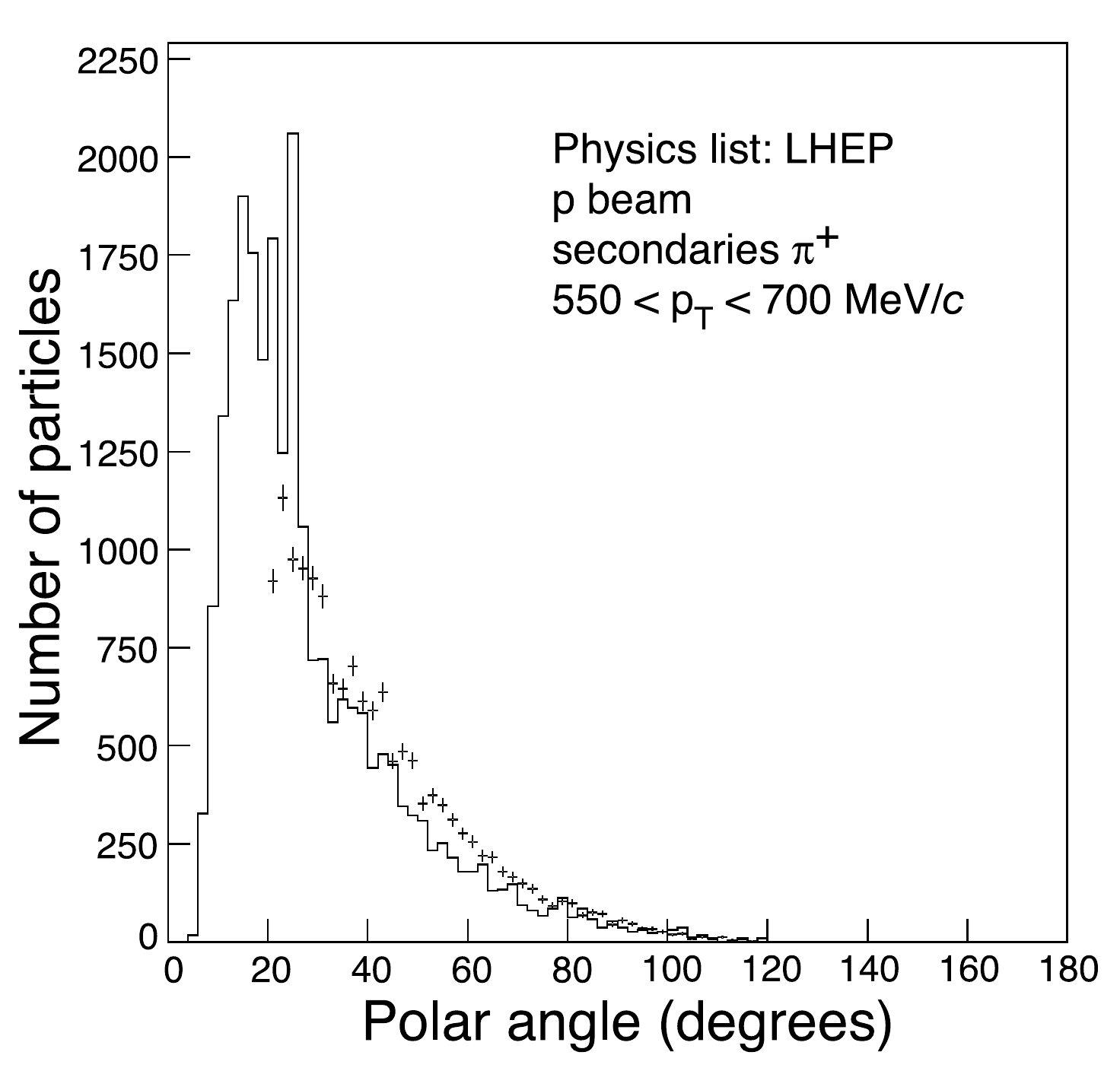} \\
\end{tabular}
\end{center}
\caption{LHEP physics list; polar-angle distributions of $\pi^+$,  
for incoming protons, in four different ranges of $p_{\rm T}$.}
\label{lheppTprotonpion}
\end{figure}

\begin{figure}[h]
\begin{center}
\begin{tabular}{cc} 
\includegraphics[width=7.5cm]{pi+_lhep_p_350to450.pdf} &
\includegraphics[width=7.5cm]{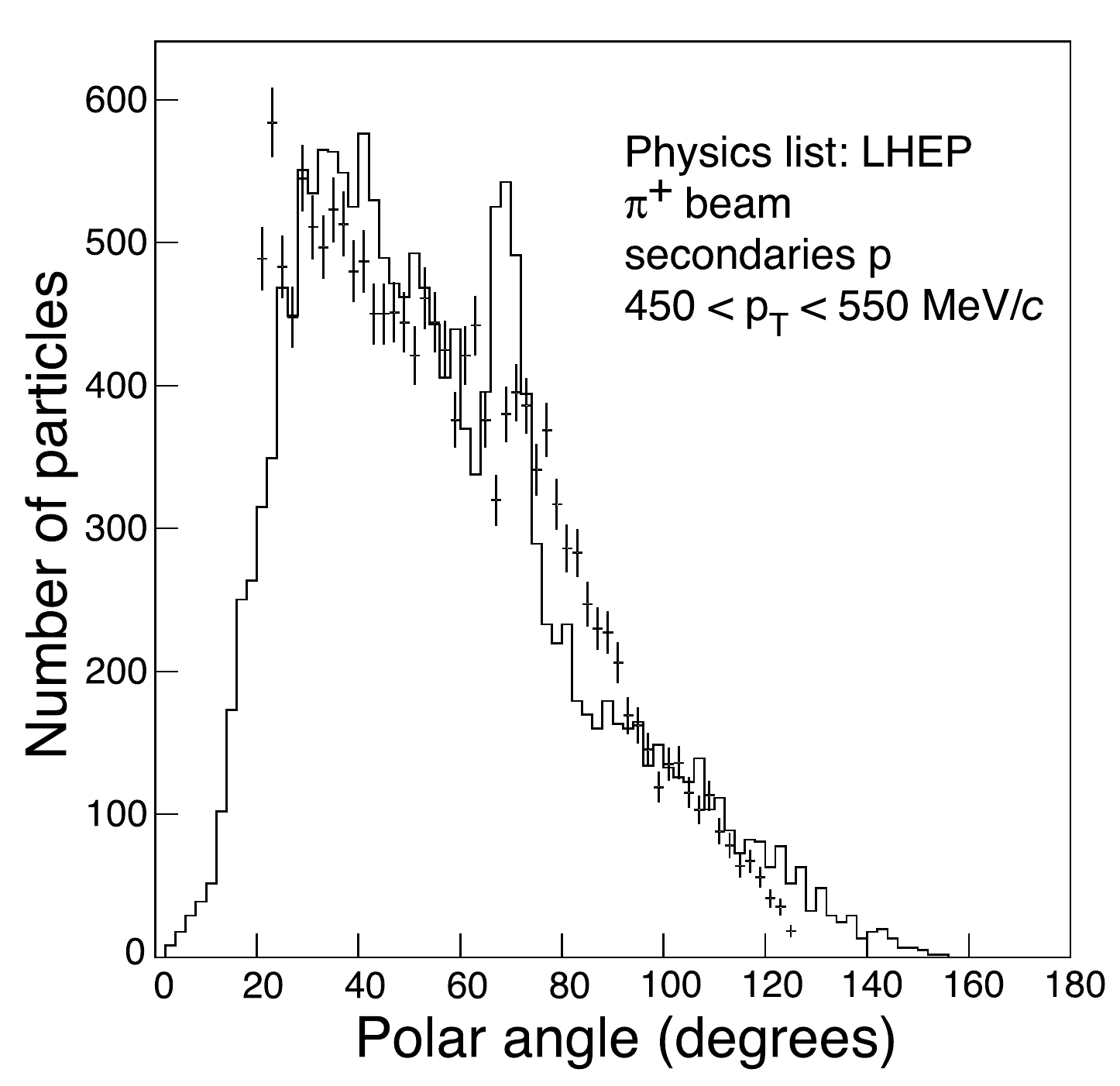} \\
\includegraphics[width=7.5cm]{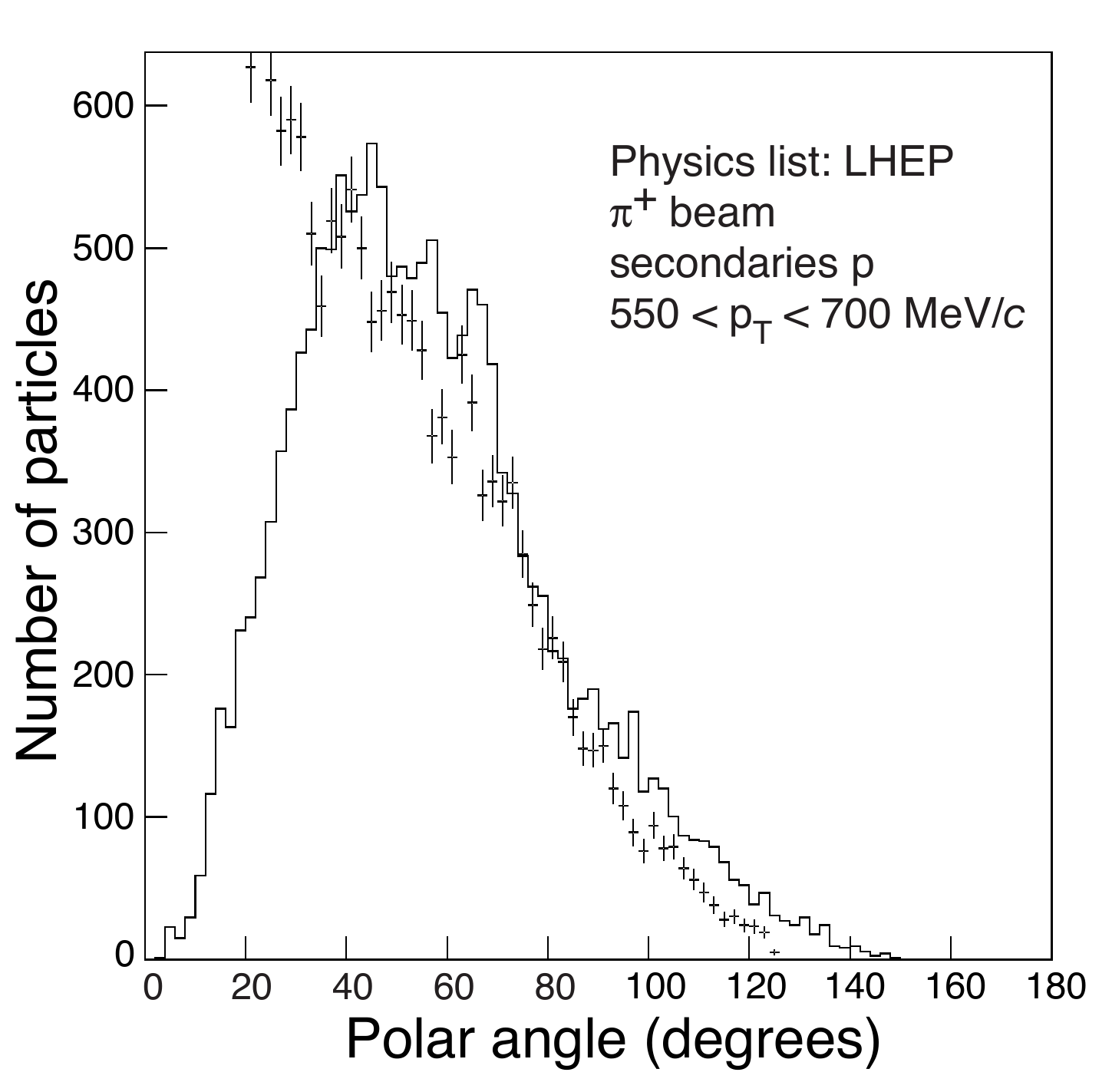} &
\includegraphics[width=7.5cm]{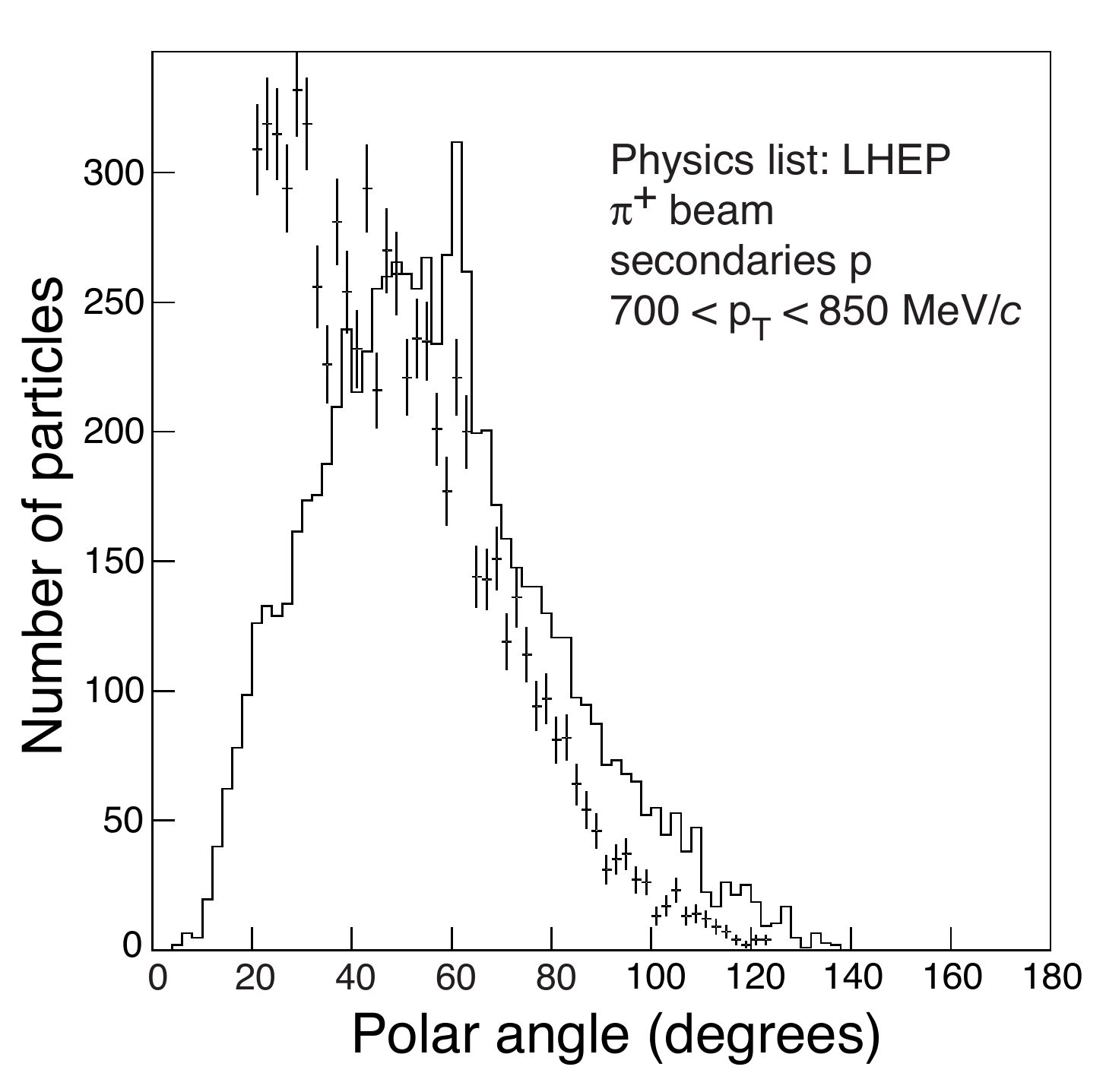} \\
\end{tabular}
\end{center}
\caption{LHEP physics list; polar-angle distributions of protons,  
for incoming $\pi^+$, in four different ranges of $p_{\rm T}$.}
\label{lheppTpionproton}
\end{figure}

Figure~\ref{lheppTprotonpion} presents for the LHEP physics list in four ranges of $p_{\rm T}$ the spectra of secondary pions from incoming protons, and Fig.~\ref{lheppTpionproton} the same for secondary protons from incoming pions. 
The conclusions from the comparison of data with simulated data are summarized in Table~\ref{tablewithconclusions}. We qualify
the various physics lists in the order `good', `acceptable', `poor', `unacceptable'. Note that our qualification refers to the restricted polar-angle range $20^\circ < \theta < 125^\circ$. 

None of the standard physics lists of Geant4 is qualified `good'.
\begin{table}[h]
\caption{Conclusions on Geant4 standard physics lists of 
hadronic interactions.} 
\label{tablewithconclusions}
\begin{center}
\begin{tabular}{|l|c|c|c|c|}
\hline
Physics list  & \multicolumn{2}{c|}{Proton beam}  
                & \multicolumn{2}{c}{$\pi^\pm$ beam} \\
\cline{2-5}
                & Secondary & Secondary & Secondary & Secondary \\
                &  protons  & $\pi^\pm$ & protons   &  $\pi^\pm$ \\
\hline
\hline
LHEP        & {\bf poor} & {\bf unacceptable} & {\bf poor} 
            & {\bf unacceptable} \\
            & (shape) & (diffr. patt.) & (el. scatt. peak 
            & (diffr. patt.) \\
            & & & in $\pi^+$ beam) &  \\
\hline
LHEP\_PRECO\_HP & & & {\bf poor} & {\bf unacceptable}  \\
            & & & (el. scatt. peak & (diffr. patt.) \\
            & & & in $\pi^+$ beam) &                \\
\hline
QGSC        & {\bf poor} & {\bf unacceptable} & {\bf poor} 
            & {\bf unacceptable} \\
            & (shape) & (diffr. patt.) & (el. scatt. peak & (diffr. patt.) \\
            & & & in $\pi^+$ beam) &          \\
\hline
QGS\_BIC    & & & {\bf poor} & {\bf unacceptable}  \\
            & & & (el. scatt. peak & (diffr. patt.) \\
            & & & in $\pi^+$ beam) &     \\
\hline
QGSP        & {\bf poor} & {\bf unacceptable} & {\bf poor} 
            & {\bf unacceptable} \\
            & (shape) & (diffr. patt.) & (el. scatt. peak 
            & (diffr. patt.) \\          
            & & & in $\pi^+$ beam) &         \\
\hline
QGSP\_BERT  & {\bf poor} & {\bf poor} & {\bf poor} & {\bf poor} \\ 
            & (shape) & (shape) & (el. scatt. peak & (shape) \\ 
            & & & in $\pi^+$ beam) &        \\  
\hline
QGSP\_BIC   & {\bf poor} & {\bf poor} & {\bf poor} & {\bf unacceptable} \\
            & (shape) & (shape) & (el. scatt. peak 
            & (diffr. patt.) \\          
            & & & in $\pi^+$ beam) &     \\ 
\hline
QBBC        & {\bf unacceptable} & {\bf acceptable} & {\bf unacceptable} 
            & {\bf acceptable} \\
            & (el. scatt. peak) & & (el. scatt. peak) &         \\
\hline
FTFC        & & & {\bf unacceptable} & {\bf acceptable} \\
            & & & (el. scatt. peak) &          \\
\hline
FTFP        & {\bf unacceptable} & {\bf poor} & {\bf unacceptable} 
            & {\bf acceptable} \\
            & (el. scatt. peak) & (shape) & (el. scatt. peak) &         \\
\hline
FTFP\_BERT  & {\bf unacceptable}  & {\bf acceptable} & {\bf unacceptable} 
            & {\bf acceptable} \\
            & (el. scatt. peak)  &  & (el. scatt. peak) & \\
\hline
\end{tabular}
\end{center}
\end{table}

The unphysical peak in the distribution of secondary protons
around $70^\circ$ is consistent with the kinematics of elastic
scattering of the incoming particle with a proton at rest.
The diffraction-like pattern in the distribution of seconday pions
is consistent with diffractive scattering of the incoming
particle on a stationary disc with the diameter of a nucleon.  
We conjecture that the differences between data and simulated 
Geant4 data arise dominantly from an inadequate description of 
the elastic scattering and diffraction scattering of incoming 
beam particles on nucleons embedded in a nucleus. 

For the analysis of our data, we have used for incoming beam protons 
the QGSP\_BIC physics list and see no strong reason to reconsider 
this choice. For incoming beam pions, none of the standard 
physics lists for hadronic interactions was acceptable, so we 
had to build our private HARP\_CDP physics list. 
This physics list starts from the QBBC physics list 
(see Table~\ref{physicslistsoverview}). Yet  
the Quark--Gluon String Model is replaced by the 
FRITIOF string fragmentation model for
kinetic energy $E>6$~GeV; for $E<6$~GeV, the Bertini 
Cascade is used for pions, and the Binary Cascade for protons; 
elastic and quasi-elastic scattering is disabled\footnote{Our
Monte Carlo simulation will be discussed in the necessary 
detail in a forthcoming paper.}. 

We qualify our HARP\_CDP physics list as `acceptable' in 
all four categories listed in 
Table~\ref{tablewithconclusions}.

\section{Summary}

We have presented significant disagreements in the laboratory polar-angle distributions between data from the HARP large-angle spectrometer, and data simulated by various physics lists in the Geant4 simulation tool kit. 
An unphysical peak for secondary protons near $\theta = 70^\circ$, and an unphysical diffraction-like pattern for secondary pions appear as dominant problems.

\clearpage

\section*{Acknowledgements}

We are indebted to J.~Apostolakis, G. Folger, A. Ribon and D.~Wright for a useful discussion on Geant4 intricacies. 
We are greatly indebted to many technical collaborators whose 
diligent and hard work made the HARP detector a well-working 
instrument. We thank all HARP colleagues who devoted time and 
effort to the design and construction of the detector, to data taking, 
and to setting up the computing and software infrastructure. 
We express our sincere gratitude to HARP's funding agencies 
for their support.  


\end{document}